\documentclass[acmsmall]{acmart}

\setcopyright{cc}
\setcctype{by}
\acmJournal{PACMSE}
\acmYear{2026} \acmVolume{3} \acmNumber{FSE} \acmArticle{FSE162}
\acmMonth{7} \acmDOI{10.1145/3808169}

\usepackage{adjustbox}
\usepackage{amsmath,amssymb,amsfonts}
\usepackage[linesnumbered,ruled,vlined]{algorithm2e}
\usepackage{setspace}
\usepackage{algorithmic}
\usepackage{float}
\usepackage{graphicx}
\usepackage{textcomp}
\usepackage{xcolor}
\usepackage{multirow}
\usepackage{mdframed}
\usepackage{balance}
\usepackage{booktabs}
\usepackage{tcolorbox}
\usepackage{makecell}
\usepackage{threeparttable}
\usepackage{colortbl}
\usepackage{subfigure}
\usepackage{hhline}
\usepackage{pifont}
\usepackage{listings}
\usepackage{enumitem}
\usepackage{caption}
\usepackage{wrapfig}

\usepackage[dvipsnames]{xcolor} 
\usepackage{tikz}

\newcommand{\circlednumber}[2][]{\tikz[baseline=(char.base)]{
    \node[shape=circle, 
          fill=orange,             
          inner sep=1pt,             
          minimum size=1em,        
          text=black,              
          font=\sffamily,          
          #1] (char) {#2};}}

\newcommand{\zq}[1]{\textcolor{black}{\textbf{}#1}}
\newcolumntype{Z}{>{\zq\bgroup}c<{\egroup}}

\definecolor{ggray}{HTML}{eff0f0}
\definecolor{gggray}{HTML}{E8E8E8}
\definecolor{ggggray}{HTML}{BEBEBE}

\newcommand{\ie}{\textit{i.e.,}}
\newcommand{\eg}{\textit{e.g.,}}

\newcommand{\kg}{C-Rust Pointer KG}

\newcommand{\ourtool}{\textsc{PtrTrans}}
\newcommand{\crust}{\textsc{c2rust}}
\newcommand{\crown}{\textsc{Crown}}
\newcommand{\pr}{\textsc{PR}\textsuperscript{2}}
\newcommand{\flour}{\textsc{FLOURINE}}
\newcommand{\Syzygy}{\textsc{Syzygy}}
\newcommand{\RustMap}{\textsc{RustMap}}

\newcommand{\ourtoolPA}{%
  $\textsc{PtrTrans}_{\text{PS}}$%
}
\newcommand{\ourtoolRA}{%
  $\textsc{PtrTrans}_{\text{RA}}$%
}

\newcommand{\ourtoolPU}{%
  $\textsc{PtrTrans}_{\text{PU}}$%
}

\newcommand{\ourtoolError}{%
  $\textsc{PtrTrans}_{\text{EC}}$%
}

\newcommand{\claude}{Claude-3.7-Sonnet}
\newcommand{\gemini}{Gemini-3-Pro}

\usepackage[utf8]{inputenc}
\usepackage[english]{babel}

\newcounter{finding}

\newcommand{\distance}{5pt}
\begin{abstract}

Translating C code into safe Rust is an effective way to ensure memory safety. Compared to rule-based approaches, which often produce largely unsafe Rust code, LLM-based methods generate more idiomatic and safer Rust by leveraging extensive training on human-written code. Despite their promise, existing LLM-based approaches still struggle with project-level C-to-Rust translation. They typically partition a C project into smaller units (\eg{} functions) based on call graphs and translate them in a bottom-up manner to resolve dependencies. However, this unit-by-unit paradigm often fails to handle pointers due to the lack of a global view of their usage.
To address this limitation, we propose a novel C-to-Rust Pointer Knowledge Graph (KG) that augments code dependency graphs with two types of pointer semantics: (i) pointer usage information, which captures global behaviors such as points-to flows and lifts low-level struct interactions to higher-level abstractions; and (ii) Rust-oriented annotations, which encode ownership, mutability, nullability, and lifetime.
Building on this KG, we further propose \ourtool{}, a project-level C-to-Rust translation approach. In \ourtool{}, the KG provides LLMs with comprehensive global pointer semantics, guiding them to generate safe and idiomatic Rust code.
Experimental results show that \ourtool{} reduces unsafe usages in translated Rust by 99.9\% compared to both rule-based and conventional LLM-based methods, while achieving 29.3\% higher functional correctness than fuzzing-enhanced LLM approaches.
\end{abstract}

\keywords{C-to-Rust Translation, Large Language Model, Knowledge Graph}

\ccsdesc[500]{Software and its engineering~Automatic programming}

\begin{document}

\title{Project-Level C-to-Rust Translation via Pointer Knowledge Graphs}

\author{Zhiqiang Yuan}
\orcid{0000-0002-6497-9380}
\affiliation{%
  \institution{Fudan University}
  \city{Shanghai}
  \country{China}
}
\email{zhiqiangyuan23@m.fudan.edu.cn}

\author{Wenjun Mao}
\orcid{0009-0002-3305-7858}
\affiliation{%
  \institution{Fudan University}
  \city{Shanghai}
  \country{China}
}
\email{25213050301@m.fudan.edu.cn}

\author{Zhuo Chen}
\orcid{0009-0004-2676-759X}
\affiliation{%
  \institution{Fudan University}
  \city{Shanghai}
  \country{China}
}
\email{25213050006@m.fudan.edu.cn}

\author{Xiyue Shang}
\orcid{0009-0004-3778-7414}
\affiliation{%
  \institution{Fudan University}
  \city{Shanghai}
  \country{China}
}
\email{25213050322@m.fudan.edu.cn}

\author{Chong Wang}
\orcid{0000-0003-1424-6290}
\affiliation{%
  \institution{Nanyang Technological University}
  \city{Singapore}
  \country{Singapore}
}
\email{chong.wang@ntu.edu.sg}

\author{Yiling Lou}
\orcid{0000-0002-4066-3365}
\affiliation{%
  \institution{Fudan University}
  \city{Shanghai}
  \country{China}
}
\email{yilinglou@fudan.edu.cn}

\author{Xin Peng}
\orcid{0000-0003-3376-2581}
\authornote{Corresponding author: Xin Peng (pengxin@fudan.edu.cn)}
\affiliation{%
  \institution{Fudan University}
  \city{Shanghai}
  \country{China}
}
\email{pengxin@fudan.edu.cn}

\maketitle

\section{Introduction}
The C language is widely used in operating systems, embedded systems, and performance-critical applications due to its low-level control over memory and hardware~\cite{DBLP:journals/corr/abs-2201-07845,DBLP:journals/corr/PatelR13a}. However, its manual memory management and direct pointer manipulation often lead to vulnerabilities such as buffer overflows and memory leaks, posing serious threats to system stability and security~\cite{fan2020ac,hanley2023rust,DBLP:journals/corr/abs-2503-03698}.
Rust offers a modern alternative that preserves the performance characteristic of C and control while ensuring memory safety~\cite{DBLP:journals/tosem/ZhengWZCL24,DBLP:journals/tosem/XuCSZL22,DBLP:conf/sigsoft/ZhangKPX23, matsakis2014rust}. 
Translating legacy C code into Rust provides a promising path to enhance both safety and performance (\eg{} rewriting Linux kernel drivers in Rust~\cite{Rust4Linux}). 
However, manual translation is labor-intensive and prone to errors~\cite{DBLP:journals/cacm/HongR25,DBLP:journals/corr/abs-2410-24117,DBLP:conf/nips/RoziereLCL20}, making automatic code translation methods urgently needed.

Existing automatic C-to-Rust translation methods fall into rule-based and large language model (LLM)-based categories. 
Rule-based approaches rely on predefined rules to translate C code into functionally equivalent Rust~\cite{c2rust, crown, citrus, DBLP:journals/pacmpl/EmreSDH21, DBLP:journals/pacmpl/EmreBPSDH23, yin2024safemd}. 
However, such approaches often produce unreadable and unidiomatic code that fails to follow the safety features in Rust. Although some recent works attempt to rewrite this unsafe code~\cite{crown, PR2, DBLP:journals/corr/abs-2501-14257}, the resulting Rust code still contains numerous unsafe blocks, raw pointers (\texttt{*mut T}), and foreign function calls to C libraries (\texttt{extern "C" fn}), which introduce safety risks~\cite{PR2, crown, DBLP:journals/pacmpl/EmreBPSDH23,DBLP:journals/pacmpl/HongR24}.
Recently, LLM-based methods have been applied to C-to-Rust translation, which can generate more idiomatic and safer Rust code, as LLMs are trained on large corpora of human-written code~\cite{DBLP:journals/corr/abs-2501-14257, DBLP:journals/corr/abs-2405-11514, DBLP:journals/corr/abs-2404-18852}.

Although promising, existing LLM-based methods still fall short in C-to-Rust translation, especially for the challenging project-level translation. In particular, given the large scale of the codebase, existing LLM-based methods typically first partition the C project into smaller units (\eg{} \textit{Function}) based on the call graph and then translate them in a bottom-up order to ensure that dependencies are resolved in advance~\cite{DBLP:journals/corr/abs-2412-14234, DBLP:journals/corr/abs-2503-17741, DBLP:journals/corr/abs-2405-11514, DBLP:journals/corr/abs-2409-10506, DBLP:journals/ese/HongR25}. However, such a bottom-up and unit-by-unit translation paradigm often fails to handle \textit{pointers} during the C-to-Rust translation~\cite{DBLP:journals/corr/abs-2409-10506}. The core difficulty lies in the fundamental difference between pointer usage in C and Rust.

In C, pointers are highly flexible: at the \textit{definition site}, a declaration specifies only the type without constraining later operations, while at the \textit{usage site}, programmers can freely read, write, modify, or free them, with semantics dictated by runtime behavior and programmer conventions. However, in Rust, pointer usage is tightly constrained: at the \textit{definition site}, ownership, mutability, and lifetime are explicitly specified (\eg{} \texttt{\&T} for an immutable borrow, \texttt{\&mut T} for a mutable borrow), and at the \textit{usage site}, all operations must strictly adhere to these declarations, with the borrow checker enforcing consistency and preventing violations such as modifying an immutable pointer.

\begin{figure*}[htb]
    \centering
    \includegraphics[width=0.95\textwidth]{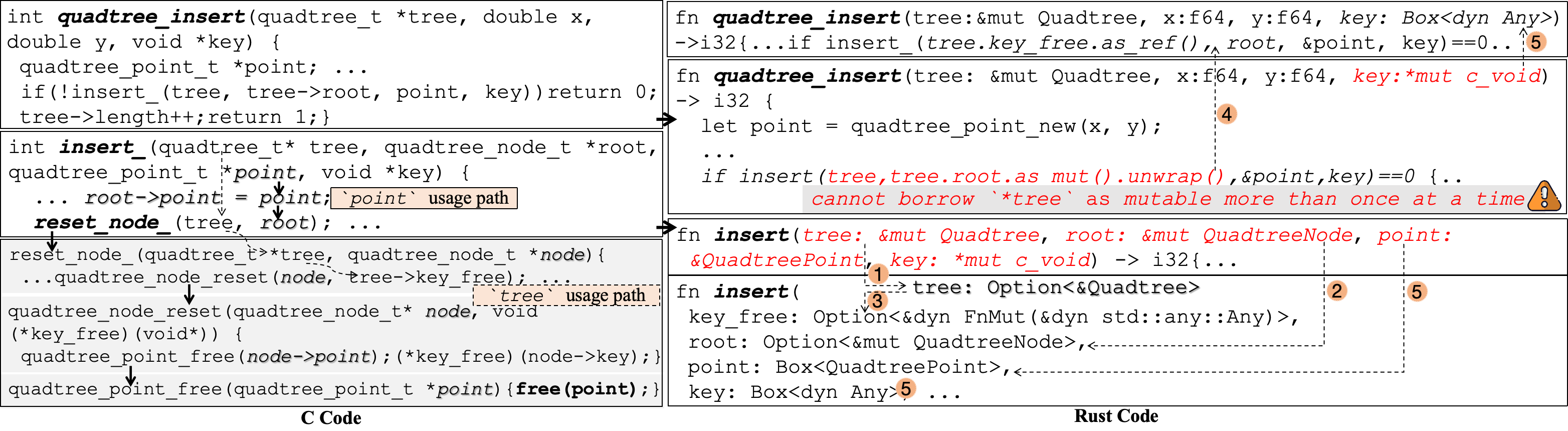}
    \caption{Motivating Example}
    \label{fig:motivate}
\end{figure*}

This difference makes bottom-up, unit-by-unit translation particularly prone to definition–use conflicts. As a single translation unit cannot reserve the usage information from higher-level call sites or across files, LLMs often misclassify pointers (\eg{} \&mut instead of an immutable borrow or Box instead of a reference) during translation. These misclassifications would cause compilation errors, such as borrow conflicts. For example, Fig.~\ref{fig:motivate} shows the \texttt{quadtree\_insert} function from Quadtree~\cite{quadtree} project translated by ChatGPT-4o~\cite{chatgpt}, producing the error ``\textit{cannot borrow \texttt{*tree} as mutable more than once at a time}''. Here, \texttt{tree} is defined as a mutable borrow (\&mut) but is borrowed mutably twice at the usage site, violating the borrowing rules in Rust.

To resolve the borrow conflict, we analyze pointer usage across the project for \texttt{quadtree\_insert} and its caller \texttt{insert\_}. 
We find that once passed into \texttt{quadtree\_insert}, \texttt{tree} never performs ownership operations (\eg{} \texttt{malloc}/\texttt{free}) and should be a borrow pointer; writes such as \texttt{tree->length++} require it to be \texttt{\&mut Quadtree}. In \texttt{insert\_}, \texttt{tree} is only used to access \texttt{key\_free} after being passed and undergoes a null check, so it should be \texttt{Option<\&Quadtree>}\circlednumber{1}, while \texttt{root} (\texttt{tree->root}) is mutably accessed and should be \texttt{Option<\&mut QuadtreeNode>}\circlednumber{2}. Yet this modification alone does not resolve the conflict, since \texttt{quadtree\_insert} still expects \texttt{tree} as mutable, creating simultaneous immutable and mutable borrows and violating Rust’s rules.
Tracing the \texttt{tree} usage path shows that its only read access is the \texttt{key\_free} in \texttt{reset\_node}. Extracting \texttt{key\_free} into a separate parameter (\ie{} \texttt{key\_free: Option<\&dyn FnMut(\&dyn Any)>})\circlednumber{3} and updating all call sites\circlednumber{4} eliminates the borrow conflict. For other pointer translations, we apply the \circlednumber{5} fix. For example, the \texttt{point} usage path reveals that it is freed in \texttt{quadtree\_point\_free}, so it should be translated as an owning pointer (\texttt{Box<QuadtreePoint>}) rather than a reference.

The motivating example above highlights the necessity of modeling the pointer-usage for both C and Rust \textit{from a global perspective}. 
First, correct Rust ownership choices (\eg{} Box vs \& vs \&mut) require project-level pointer-usage evidence and cannot be inferred reliably from a single translation unit. Second, global pointer-usage information can guide translation and enable targeted refactoring (\eg{} parameter extraction and borrow decomposition) to resolve borrow conflicts.

To this end, we first propose a novel C-Rust Pointer Knowledge Graph (KG), which models the code-unit dependencies, pointer-usage information, and Rust-oriented annotations from the project-level context. In particular, our C-Rust Pointer KG includes a code-dependency graph as its skeleton to capture relationships among code units (\eg{} \textit{Function}, \textit{Struct}, \textit{Enum}), and enriches it with two types of C–Rust pointer semantics: (1) pointer-usage information, recording global behaviors such as points-to flows, field accesses, and lower-level struct usage by higher-level units; and (2) Rust-oriented annotations, encoding ownership, mutability, nullability, and lifetimes.

Based on the \kg{} and its synergy with LLMs, we further propose \ourtool{}, a project-level C-to-Rust translation technique for translating a given C project into a safe and idiomatic Rust project. \ourtool{} leverages the constructed C-Rust Pointer KG to guide LLMs for translation. In particular, the code-dependency graph in the KG determines the translation units and their translation order, while pointer-usage information and Rust-oriented annotations provide pointer semantics for each unit, guiding LLMs in the translation process. \ourtool{} employs an incremental verification mechanism: after translating each unit, the generated Rust code is immediately integrated into the initialized Rust project and compiled. If compilation errors occur, \ourtool{} corrects them to prevent error accumulation. 
Although both existing LLM-based \zq{techniques~\cite{DBLP:journals/corr/abs-2412-14234, DBLP:journals/corr/abs-2503-17741, DBLP:journals/corr/abs-2409-10506, DBLP:journals/ese/HongR25}} and our approach adopt a compositional (\ie{} unit-by-unit) translation paradigm, our approach differs substantially by incorporating our C–Rust Pointer KG, which effectively preserves global pointer information. Such information, overlooked in prior techniques, enhances translation accuracy and reduces repeated trial-and-error during error correction.

We evaluate \ourtool{} on 16 real-world C projects (up to 14,829 lines of code). First, we assess the idiomaticity of Rust projects generated by \ourtool{}. 
\zq{Results show that \ourtool{} reduces Lint warnings by 94.9\% (from 6,802 to 349) and unsafe code usage by 99.9\% (from 141,866 to 85) compared with \crown{}\cite{crown}, and reduces Lint warnings by 91.6\% (from 4,135 to 349) and unsafe code usage by 99.9\% (from 134,185 to 85) compared with \pr{}\cite{PR2}, demonstrating stronger idiomaticity and safety.}
\zq{Second, we evaluate translation correctness and find that \ourtool{} improves the average functional equivalence rate by 29.3\% over the fuzzing-enhanced LLM-based method \flour{}~\cite{DBLP:journals/corr/abs-2405-11514} on projects with $\text{LoC}<4\text{K}$, and by 53.7\% on larger projects with $4\text{K}<\text{LoC}<15\text{K}$.}
Third, we perform an ablation study, demonstrating that pointer-usage information and Rust-oriented annotations are essential for enhancing the translation performance of \ourtool{}. Fourth, we analyze the impact of factors such as lines of code, function dependencies, and pointer counts. Results show \ourtool{} mitigates challenges from these complexities more effectively than \flour{}. 
\zq{Finally, we demonstrate that \ourtool{} generalizes well across different LLMs, leading to improved translation performance.}

In summary, this paper makes the following contributions:
\begin{itemize}[topsep=3pt, leftmargin = 13pt]

    \item We design a novel \kg{}, which comprehensively models code-unit dependencies, pointer-usage information, and Rust-oriented annotations from the project-level context.
    
    \item We propose a novel project-level C-to-Rust translation technique \ourtool{}, which leverages \kg{} and synergy with LLMs to translate C projects into safe and idiomatic Rust.
         
    \item We conduct a comprehensive evaluation and show that \ourtool{} generates more idiomatic, safer Rust code with higher correctness than existing LLM-based methods. Ablation study showing that pointer-usage information and Rust-oriented annotations are crucial for improving \ourtool{} translation performance.

\end{itemize}

\section{Related Work}~\label{sec:related}
Code translation has been extensively studied in existing literature\cite{DBLP:conf/icse/PanIKSWMSPSJ24, DBLP:journals/pacmse/Yang0YK0LHMJ024, DBLP:journals/corr/abs-2410-24117, guan2025repotransagentmultiagentllmframework, DBLP:journals/corr/abs-2412-08035, DBLP:conf/nips/RoziereLCL20,10938485}. Given the  substantial differences among programming languages, code translation techniques are typically tailored to specific language pairs (\eg{} C-to-Rust, Java-to-Python, Python-to-C). Therefore, in this section, we focus on the existing work on C-to-Rust translation. Translating C programs into Rust has received significant attention due to the memory safety and performance characteristics of Rust. Current translation approaches can be categorized into rule-based and LLM-based methods.

\textit{Rule-based translation methods} rely on  manually crafted transformation rules to convert C code into Rust~\cite{DBLP:journals/corr/abs-2412-15042,DBLP:journals/ieeesp/Larsen24} (\eg{} Corrode~\cite{Corrode} and \crust{}~\cite{c2rust}).
However, such methods often produce Rust code that is unsafe.
To reduce unsafety, recent studies rewrite rule-generated Rust code via handcrafted rules~\cite{DBLP:conf/icse/LingYWWCH22, DBLP:journals/pacmpl/EmreBPSDH23, DBLP:journals/pacmpl/EmreSDH21,crown, DBLP:journals/pacmpl/HongR24, DBLP:conf/kbse/HongR24} or LLMs~\cite{DBLP:journals/corr/abs-2506-01427, PR2, DBLP:journals/corr/abs-2404-18852,DBLP:journals/corr/abs-2504-15254, DBLP:journals/corr/abs-2501-14257}.
For example, \crown{}~\cite{crown} refactors unsafe Rust pointers into safe ones by analyzing ownership, while \pr{}~\cite{PR2} refines \crust{}~\cite{c2rust}-generated Rust by rewriting function bodies with LLMs.
Nevertheless, the generated Rust code still contains numerous unsafe code blocks and exhibits non-idiomatic Rust patterns~\cite{DBLP:conf/ndss/LiWLSK25}.

\textit{LLM-based translation methods} typically generate more readable and idiomatic Rust code due to training on large corpus of human-written code~\cite{11126570, DBLP:journals/corr/abs-2501-14257, DBLP:journals/corr/abs-2405-11514, DBLP:journals/corr/abs-2404-18852, DBLP:journals/corr/abs-2503-12511, DBLP:journals/corr/abs-2505-15858, shiraishitoward, DBLP:journals/corr/abs-2410-24117}. 
Currently, LLMs demonstrate strong capabilities in code generation~\cite{DBLP:journals/corr/abs-2308-01240, DBLP:journals/tosem/DongJJL24, DBLP:journals/pacmse/Yuan0DW00L24}, program repair~\cite{DBLP:conf/icse/FanGMRT23, DBLP:conf/icse/XiaWZ23, DBLP:conf/icse/BouzeniaDP25}, and code translation~\cite{DBLP:journals/corr/abs-2409-19894, DBLP:journals/pacmse/Yang0YK0LHMJ024, DBLP:journals/corr/abs-2410-24117}, which have substantially advanced the state of LLM-based C-to-Rust translation.
From the perspective of translation strategy, existing LLM-based C-to-Rust approaches can be divided into (i) all-at-once and (ii) unit-by-unit translation.

All-at-once translation refers to submitting all C code units to LLMs in a single batch to translate into the corresponding Rust code~\cite{DBLP:conf/icse/PanIKSWMSPSJ24, DBLP:journals/corr/abs-2503-18305, DBLP:journals/corr/abs-2411-13990, DBLP:journals/corr/abs-2505-10708, DBLP:journals/corr/abs-2405-18574}.
For example, SPECTRA~\cite{DBLP:journals/corr/abs-2405-18574} generates specifications for the functions and then feeds them back to LLMs to enhance translation. \flour{}~\cite{DBLP:journals/corr/abs-2405-11514} uses fuzzing to check the equivalence of translated Rust code and applies test-enhanced method to repair non-equivalent code. However, such methods are effective only for relatively small C programs (\eg{} up to 600 lines of code)~\cite{DBLP:journals/corr/abs-2505-10708, DBLP:conf/icse/PanIKSWMSPSJ24, DBLP:journals/corr/abs-2405-11514}.

Unit-by-unit translation partitions a project into independent units based on the call graph and translating them in a bottom-up order~\cite{DBLP:journals/corr/abs-2412-14234, DBLP:journals/corr/abs-2503-17741, DBLP:journals/corr/abs-2409-10506, DBLP:journals/ese/HongR25}. For example, \Syzygy{}\cite{DBLP:journals/corr/abs-2412-14234} uses test cases to improve LLM translation performance. However, these methods lack global pointer semantic awareness during translation. As a result, lower-level units cannot capture pointer usage in higher-level contexts, making it difficult to accurately translate units such as \textit{structs} and \textit{unions} without manual intervention\cite{DBLP:journals/corr/abs-2503-17741, DBLP:journals/corr/abs-2412-14234}.

Different from existing C-to-Rust translation approaches, \ourtool{} relies on our \kg{} and its synergy with LLMs. 
Our KG not only captures dependency relationships among code units in the C project but also integrates two forms of pointer knowledge: pointer-usage information and Rust-oriented annotations.
Pointer-usage information captures global pointer behavior and provides missing context during bottom-up translation (\eg{} how lower-level structs or unions are used in higher-level functions).
Explicit Rust-oriented annotations (\eg{} ownership and mutability) serve as factual constraints to guide generation and error repair, reducing reliance on implicit inference from limited context and improving translation accuracy.

\section{Approach}
As shown in Fig.~\ref{fig:pipeline}, \ourtool{} comprises three stages: \kg{} construction (Sec~\ref{sec:app:kg}), KG-guided translation (Sec~\ref{sec:app:translation}), and KG-guided error correction (Sec~\ref{sec:error_fix}). 
Given a C project, \ourtool{} first constructs a \kg{} using a code-dependency graph as its skeleton to capture dependencies among project code units, enriched with pointer global usage information and Rust-oriented annotations. 
During KG-guided translation, the dependency graph determines the translation units and their bottom-up translation order, while pointer usage and annotations supply essential semantic context for generating safe Rust code.
Subsequently, during KG-guided correction, \ourtool{} incrementally integrates each translated Rust unit into the initialized Rust project for syntax validation, and any detected error is immediately corrected before proceeding. Such an incremental strategy enables timely error resolution and prevents error accumulation. 

In particular, our approach has two novelties. (1) \textbf{\kg{}}: we design a novel \kg{}, which can comprehensively model the code-unit dependencies, pointer-usage information, and Rust-oriented annotations from the project-level context; (2) \textbf{KG-guided code translation}: we propose a novel project-level C-to-Rust translation paradigm guided by the \kg{} and its synergy with LLMs.

\subsection{C-Rust Pointer Knowledge Graph Construction}\label{sec:app:kg}
In this phase, our approach constructs a \kg{} for the given C project. In addition to a basic \textit{code-dependency graph} which preserves relationships among code units, our \kg{} further incorporates two categories of C–Rust pointer semantics: (i) \textit{pointer-usage information} which records global pointer behaviors,  and (ii) \textit{Rust-oriented annotations} which encode explicit Rust-oriented pointer labels. In the following, we first introduce the schema of our \kg{} and then describe the detailed construction process.

\subsubsection{Schema of Pointer KG}

Our \kg{} builds on a code-dependency graph as the structural backbone, and is further enriched with C–Rust pointer semantics, including pointer-usage information and Rust-oriented annotations.
In particular, our KG focuses on the pointers in function parameters, function return values, and members of structs and unions, as these pointers define module interfaces, which are crucial for inter-module semantic consistency.

\begin{figure*}[htb]
    \centering
    \includegraphics[width=0.85\textwidth]{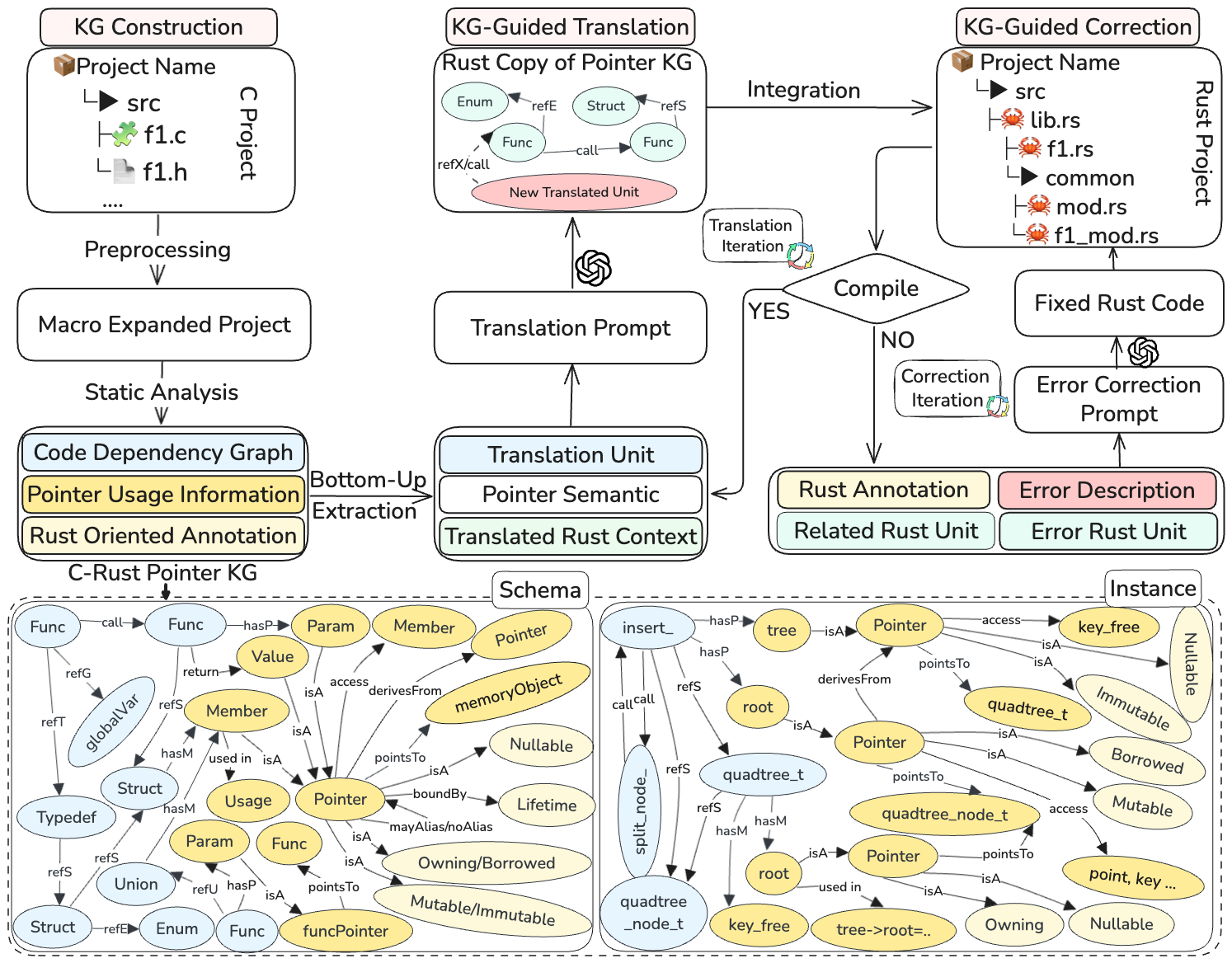}
    \caption{Workflow of \ourtool{}}
    \label{fig:pipeline}
\end{figure*}

\textbf{Code-Dependency Graph} depicts dependency relationships between code units in the project. 

\begin{itemize}[leftmargin=10pt, topsep=5pt]

\item \textit{Code Units.} Each function (\textbf{Func}), struct (\textbf{Struct}), enum (\textbf{Enum}), union (\textbf{Union}), typedef (\textbf{Typedef}), and global variable (\textbf{globalVar}) is included as a code unit entity (\ie{} nodes) in KG. 
\zq{The code unit definition follows the common granularity adopted in existing C-to-Rust translation works~\cite{guan2025repotransagentmultiagentllmframework, DBLP:journals/corr/abs-2412-14234,DBLP:journals/corr/abs-2503-17741}.}
Each code unit entity is linked to its corresponding C  implementation and annotated with metadata such as its source file name and type (\ie{} \texttt{.c} or \texttt{.h} file).

\item \textit{Dependencies.} The reference and invocation relationships between code units are included as relations (\ie{} edges) in KG.  They include \textit{Func}-to-\textit{Func} call relationships (\textbf{call}) and  a variety of reference types (\textbf{refX}), such as \textit{Func}-to-\textit{Struct} (refS), \textit{Struct}-to-\textit{Enum} (refE), and \textit{Struct}-to-\textit{Union} (refU). 
\zq{The dependency relations are defined following those commonly used in compiler intermediate representations and static-analysis frameworks, and collectively capture the key connections among code units.}

\end{itemize}

\textbf{Pointer-Usage Information} captures project-wide pointer behaviors, including inter-module interactions that cannot be inferred within a single module. This global perspective could reserve  comprehensive semantic information beyond local scopes.

\begin{itemize}[leftmargin=10pt, topsep=5pt]

\item  Represented by the entities: \textbf{Param}, denoting function input parameters; \textbf{Value}, representing function return values; \textbf{Member}, referring to fields within structs or unions; \textbf{Pointer}, indicating whether a \textit{Param}, \textit{Value}, or \textit{Member} is of pointer type; 
\textbf{memoryObject} refers to a concrete memory instance that a pointer may reference (\eg{} an instance of a struct, union, or array);
\textbf{funcPointer}, representing pointers that specifically point to functions; and \textbf{Usage}, capturing the usage context of struct or union member pointers within the project.

\item Represented by the relations: 
\textbf{derivesFrom} indicates whether the value of a pointer variable is obtained by dereferencing another pointer and accessing one of its fields, representing data-flow dependencies between pointers.
\textbf{mayAlias/noAlias} characterize whether two pointers may refer to the same object across all calling contexts; \textbf{pointsTo} records all possible memory instances a pointer may reference; \textbf{Access} represents member fields accessed after a struct or union parameter is passed into a function; and \textbf{usedIn} captures the usage scenarios of struct or union member pointers within the project.

\end{itemize}

\textbf{Rust-Oriented Annotations} capture Rust-oriented semantics of pointers by analyzing their lifecycles across the entire project. These annotations (including ownership, mutability, nullability, and lifetimes) cannot be reliably inferred from a single translation unit, as they require project-wide context to determine how pointers are defined, used, and propagated.

\begin{itemize}[leftmargin=10pt, topsep=5pt]

\item  This annotation is represented by the following entities.
\textbf{Owning/Borrowed} distinguishes whether a pointer implies ownership of a resource or merely a borrow, guiding its representation as an owning type (\eg{} \texttt{Box}, \texttt{Vec}) or a borrowed type (\eg{} \texttt{\&T}). \textbf{Mutable/Immutable} indicates whether the data it points to can be modified, dictating the use of either an immutable reference (\texttt{\&T}) or a mutable reference (\texttt{\&mut T}) in Rust.
\textbf{Lifetime} denotes a compiler-enforced safety contract ensuring a reference cannot outlive its borrowed data, specified either explicitly with generic annotation (\eg{} \texttt{'a}), static annotation (\eg{} \texttt{'static}), or implicitly through compiler elision mechanism (\texttt{'\_}).
\textbf{Nullable} indicates that pointers in C are, by default, permitted to be \texttt{NULL}, reflecting the principle of ``fail-safe by default~\cite{DBLP:journals/pieee/SaltzerS75}''. Since the C type system does not enforce non-nullability, any pointer may theoretically hold a null value. Consequently, when translating C pointers to Rust, nullable pointers must be wrapped with \texttt{Option<T>} to ensure safety. 
Although this default permissiveness may introduce redundant null checks in the generated Rust code, the risk of null pointer dereferencing is eliminated~\cite{c2rust, crown}.

\end{itemize}

Fig.~\ref{fig:pipeline} shows the schema of \kg{}, illustrating the types of entities and relations. It also showcases a partial KG instance for the Quadtree~\cite{quadtree} project. The schema design of our KG aims at capturing global pointer semantics, which can subsequently provide LLMs with cross-module context for more accurate Rust translation.
For instance, for the failed translation in Fig.~\ref{fig:motivate}, the \texttt{insert\_} parameters \texttt{tree} and \texttt{root} are now annotated in the KG as an immutable reference (\&) and a mutable reference (\&mut), respectively, enabling LLMs to assign correct mutability. 
Additionally, the \textbf{derivesFrom} relation indicates that \texttt{root} is derived from the structure pointed to by \texttt{tree} (\ie{} quadtree\_t). After being passed to \texttt{insert\_}, \texttt{tree} only reads its \texttt{key\_free} member, while \texttt{root} accesses all its members (\eg{} \texttt{point} and \texttt{key}). By leveraging this global usage information, LLMs extract \texttt{key\_free} from \texttt{quadtree\_t} as a separate function parameter, avoiding borrow conflicts and improving both the correctness and safety of the Rust translation.

\subsubsection{Code-Dependency Graph Extraction}
For the given C project, we first construct a code-dependency graph by capturing all dependencies among various code units. 
The construction process comprises two phases: preprocessing and dependency extraction. During preprocessing, \ourtool{} follows \zq{rule-based methods~\cite{c2rust, crown, DBLP:journals/pacmpl/EmreSDH21} to perform macro expansion on the C project}, eliminating the complexity introduced by macro definitions and invocations. During dependency extraction, \ourtool{} uses static analysis to extract different types of code units (\eg{} \textit{Func}, \textit{Struct}) and their dependencies within the macro-expanded C project. 
Each unit is annotated with corresponding code implementation and file information (\ie{} whether from a \texttt{.c} or \texttt{.h} file before macro expansion). 
Notably, macro expansion introduces extensive system header code into \texttt{.c} files, which \ourtool{} filters out during dependency extraction to focus on project source code.

\subsubsection{Pointer-Usage Information Extraction}
We then extend the dependency graph above with pointer-usage information, which is mainly extracted for two types of code units: \textit{function-type units} (\eg{} pointers in function parameters) and \textit{struct/union-type units} (\eg{} pointer members).  We perform a context-sensitive and field-sensitive analysis to distinguish pointer behaviors across different calling contexts and struct/union members.  Guided by the code-dependency graph, we adopt a bottom-up analysis strategy that starts from leaf units and moves upward, propagating pointer semantics to reduce redundant analysis. We then describe how to extract pointer-usage information from the C project in detail.

\begin{figure}[htb]
    \centering
    \includegraphics[width=0.8\textwidth]{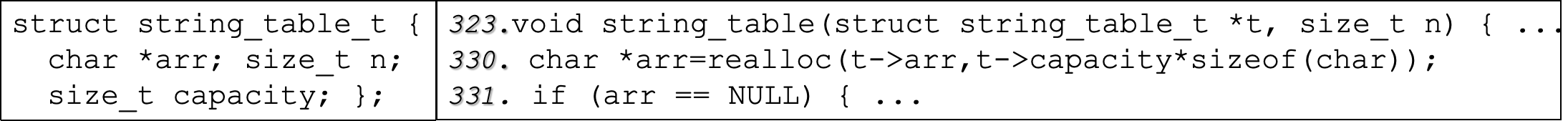}
    \caption{Definition and Usage Context of \texttt{string\_table\_t}~\cite{libtree} Struct}
    \label{fig:struct}
\end{figure}

For function-type code units, pointer analysis applies two complementary forms of inter-procedural analysis to collect global information. 
First, to obtain \textbf{Access} relations for function pointer parameters, we perform a downward analysis. Specifically, we scan the current function for struct/union member accesses and, crucially, recursively traverse any callees to which the pointer is passed. This strategy aggregates a complete set of member-access information across the entire call chain originating from the function.
Second, to identify \textbf{pointsTo}, \textbf{mayAlias/noAlias}, and \textbf{derivesFrom} relations, we conduct an upward inter-procedural analysis. For each pointer parameter, we collect its set of possible referenced objects (\textbf{memoryObjects}) by examining all direct call sites of the function, which directly defines \textbf{pointsTo}. 
Based on the collected \textbf{memoryObjects} sets, we perform pointer alias analysis: if the \textbf{memoryObjects} sets of two pointer parameters intersect, they are classified as potential aliases (\textbf{mayAlias}); otherwise, they are \textbf{noAlias}. The \textbf{derivesFrom} relation is determined by examining the actual arguments passed to a function at each direct call site. If one argument is derived from the struct/union of another argument (\eg{} \&s and \&s->field), we establish a \textbf{derivesFrom} relation between the corresponding formal parameters.

For struct/union-type units,  we traverse each pointer member and perform static analysis on the source C project to identify all code lines related to operations on that member, constructing its specific \textbf{Usage}. For example, as illustrated in Fig.~\ref{fig:struct}, for the member pointer \texttt{arr} in the \texttt{string\_table\_t} struct, we extract key code lines involving \texttt{arr} from \texttt{string\_table} (Lines 330–331). The \texttt{realloc} operation in \textbf{Usage} enables LLMs to accurately translate \texttt{arr} as a dynamically sized array (\ie{} \texttt{Vec<u8>}), rather than a fixed-size array (\ie{} \texttt{[u8; N]}).

\subsubsection{Rust-Oriented Annotations.}
We further extend the dependency graph with Rust-oriented annotations, including ownership, mutability, nullability (defaulting to \texttt{NULL}), and lifetimes, which are extracted with the following set of our predefined rules.

\textit{Ownership Annotation.}
In C, pointer ownership can be understood as a resource management contract, typically characterized by how a pointer manages memory. To distinguish between owning and borrowing pointers in static analysis, we establish specific determination rules focusing on three critical stages in the lifecycle of a pointer: \textit{creation}, \textit{destruction}, and \textit{transfer}. A pointer satisfying any of the following criteria is classified as owning; otherwise, it is considered borrowing.

\begin{itemize}[leftmargin=10pt, topsep=5pt]
\item  \textit{Rule 1: Creation Stage.} A pointer is deemed owning if it acquires memory through allocation functions such as \texttt{malloc}, \texttt{calloc}, or \texttt{realloc}. For example, as illustrated in Fig.~\ref{fig:struct}, the member pointer \texttt{arr} of the \texttt{string\_table\_t} struct obtains memory via \texttt{realloc}.

\item  \textit{Rule 2: Destruction Stage.}
If the code associated with the pointer eventually invokes the corresponding deallocation function (\eg{} \texttt{free}) to release the memory pointed to, the pointer is regarded as owning. For instance, in Fig.~\ref{fig:motivate}, the pointer parameter \texttt{point} in the \texttt{quadtree\_point\_free} function is freed using \texttt{free}.

\item  \textit{Rule 3: Transfer Stage.}
A pointer is classified as owning if it is stored in a new data structure (\eg{} as a struct member or container element) and the data structure later assumes responsibility for its release.
For example, in the \texttt{insert\_} function, the parameter \texttt{point} is assigned with \texttt{root->point = point}. This assignment counts as an ownership transfer only because the corresponding cleanup function (\eg{} \texttt{quadtree\_free\_node}) explicitly calls \texttt{free(node->point)}, showing that the data structure assumes responsibility for deallocation.

\end{itemize}

\textit{Mutability Annotation.}
In C, mutability is determined by observing the behavior of pointer throughout its accessible scope. A pointer is classified as a mutable reference (\&mut T) if any explicit or indirect write operation is performed on the memory it references; otherwise, it is considered an immutable reference (\&T).

\begin{itemize}[leftmargin=10pt, topsep=5pt]
\item  \zq{\textit{Rule 1: Code Behavior.}}
A pointer is considered mutable if it is involved in any of the following write behaviors:
(1) Direct Modification: The pointer directly alters the memory it references, such as through assignment (\eg{} \texttt{*p=..}), struct/union member assignment (\eg{} \texttt{p->member = ...}), or modification via unary operators (\eg{} \texttt{p->member++}, \texttt{(*p)-{}-}); (2) Transitive Modification via Standard Functions: The pointer is passed as a mutable argument to standard library functions that modify memory, such as \texttt{strcpy}, \texttt{memset} or \texttt{memcpy};
(3) Modification via Aliasing and Type Casting: An alias of the pointer is created, potentially by casting away a \texttt{const} qualifier (\texttt{(char*)p}), and this alias is then used to modify the original memory.
\end{itemize}

\textit{Lifetime Annotation.}
In C, the lifetime refers to the validity of a pointer from allocation to deallocation and is implicit. Lifetime annotations make these constraints explicit for the Rust compiler, either through explicit markers (\eg{} generic \texttt{'a} or the special \texttt{'static}) or implicitly via lifetime elision mechanism (\texttt{'\_}).  We establish rules to determine whether function signatures and pointer members in structs or unions require explicit lifetime annotations or can be elided.

\begin{itemize}[leftmargin=10pt, topsep=5pt]

\item  \textit{Rule 1: Lifetime Annotation for Function Signatures.}
For function signatures, explicit lifetime annotations are required when the Rust compiler cannot unambiguously infer them. This arises in two cases. (1) Ambiguous returns: if a function returns a pointer that may come from multiple input parameters, a generic lifetime (\eg{} \texttt{'a}) explicitly links them. 
For example, \textit{char* func(char* s1, char* s2)} may return either input. Since the return value depends on multiple parameters, Rust requires an explicit lifetime to capture this relationship: \textit{fn func<'a>(s1: \&'a str, s2: \&'a str) -> \&'a str}.
(2) Static data: if a function returns a pointer to data with program-long validity, such as a string literal or static variable, the \texttt{'static} lifetime is annotated. When lifetimes are clear, such as a function with a single input and output reference, annotations can be omitted.

\item  \textit{Rule 2: Lifetime Annotation for Struct and Union Members.}
For borrowed pointer members in structs/unions, Rust requires a lifetime annotation. 
If it borrows from a non-static source (\eg{} a local variable), the struct/unions requires a generic lifetime parameter (\eg{} \texttt{struct MyStruct<'a>}), and the member uses that lifetime (\eg{} \texttt{member:\&'a str}). If the member consistently borrows from a static source (\eg{} string literal or global), it needs the \texttt{'static} lifetime. 

\end{itemize}

Based on the above rules, we annotate function-type units (pointer parameters and return pointers) and struct/union-type units (pointer members) within \kg{} with Rust-oriented semantics. To enable this, we first extract the usage paths of pointers at these locations.

\begin{itemize}[leftmargin=10pt, topsep=5pt]
\item \textit{For function parameters}, we perform both downward and upward analyses to capture their complete usage paths. Downward analysis tracks the parameter’s usage within the current function and in any downstream functions when it is passed as an argument. Upward analysis considers all direct call sites, capturing how the parameter is used after being passed as an actual argument to the current function. 

\item \textit{For function return pointers}, we combine internal analysis (tracking the origin of the return pointers within the function body) with call-site analysis (tracking how the return pointers are used by callers) to construct the usage path. 

\item \textit{For pointer members in structs and unions}, we perform a project-wide analysis to construct a complete usage path, tracking each member from initialization (\eg{} via \texttt{malloc} or assignment) through all accesses, modifications, and eventual deallocation (\eg{} via \texttt{free}).

\end{itemize}

\begin{figure}[htb]
    \centering
    \includegraphics[width=0.65\textwidth]{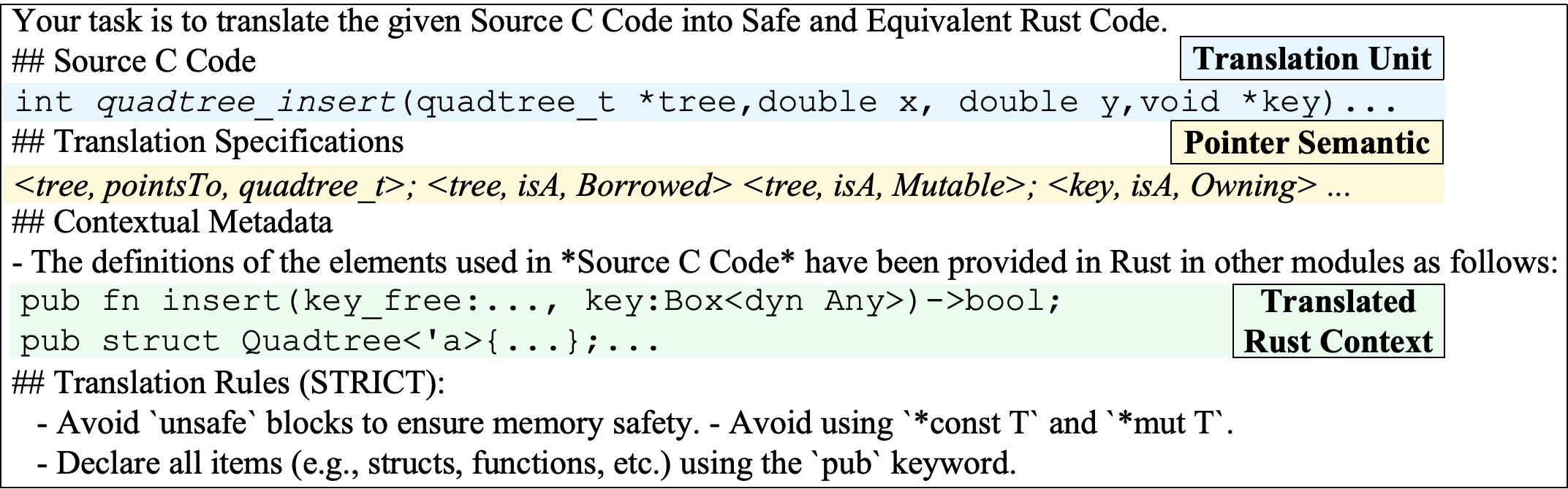}
    \caption{Code Translation Prompt}
    \label{fig:trans_prompt}
\end{figure}

The usage path is constructed via data-flow analysis, capturing the complete lifecycle of a pointer, from its creation (\eg{} via \texttt{malloc}), through intermediate uses (\eg{} writes or being passed as an argument), to destruction (\eg{} via \texttt{free}). Moreover, to reduce analysis overhead, we adopt a bottom-up strategy along the dependency graph, integrating usage paths from lower-level functions directly into the current unit to avoid redundant analysis. 
Along the extracted usage paths, we first apply the Ownership Rule to determine whether a pointer is owning. If so, the annotation is complete, as owning pointers already imply exclusive control over their values; otherwise, we further apply the Mutability and Lifetime rules. 
Lifetime annotations are introduced in function signatures and struct/union members, ensuring that Rust can correctly verify borrowing relationships.

In this process, the rules operate on usage paths and are independent of the syntactic forms of pointer operations. For pointer-related macros, we perform macro expansion prior to \kg{} construction, making pointer-related operations explicit as ordinary C statements. For example, \texttt{\#define NEW\_NODE(type) ((type*)malloc(sizeof(type)))} expands \texttt{Node *p = NEW\_NODE(Node);} into \texttt{Node *p = (Node*)malloc(sizeof(Node));}. Along the resulting usage path, the KG records \texttt{<p, pointsTo, Node>} and \texttt{<p, isA, Owning>}.
Recursive structures likewise require no special handling. For example, \texttt{struct Node\{struct Node *next;\}} defines a self-referential type, where \texttt{next} is a pointer to \texttt{Node}. When the object referenced by \texttt{next} is allocated via \texttt{malloc} and released via \texttt{free}, the KG records \texttt{<Node, hasM, next>}, \texttt{<next, isA, Owning>}, and \texttt{<next, pointsTo, Node>}. Based on these annotations, LLMs translate the recursive structure into an idiomatic Rust definition, \ie{} \texttt{struct Node\{next: Option<Box<Node>{}>\}}.


\subsection{KG-Guided Code Translation} \label{sec:app:translation}

\ourtool{} leverages the constructed \kg{} to guide LLMs to automatically translate a given C project into Rust. 
Unlike existing approaches that rely solely on LLMs to infer implicit pointer semantics knowledge in C code~\cite{DBLP:journals/corr/abs-2412-14234, DBLP:journals/corr/abs-2503-17741, DBLP:journals/corr/abs-2405-11514,DBLP:journals/corr/abs-2409-10506, DBLP:journals/corr/abs-2505-10708, DBLP:journals/ese/HongR25}, \ourtool{} leverages explicit pointer semantics from the pointer KG to assist LLMs in generating idiomatic Rust code.
Specifically, \ourtool{} first identifies translation units and determines their processing order based on the code-dependency graph.
For each translation unit, \ourtool{} extracts pointer semantic knowledge, including pointer-usage information and Rust-oriented annotations. Subsequently, \ourtool{} integrates the translation unit, the extracted pointer semantic knowledge, and the context of dependencies previously translated into Rust code into the translation prompt (see Fig.~\ref{fig:trans_prompt}). 
Based on this enriched prompt, LLMs generate the corresponding Rust code unit, which is then incrementally integrated into the Rust project,  initialized via \texttt{cargo new}, and followed by immediate compilation verification.
The translation proceeds to the next unit if no errors are detected; otherwise, an error correction mechanism is triggered (see Section~\ref{sec:error_fix}).

\subsubsection{Translation Unit Identification and Sequencing}
code-dependency graph encompasses various types of code units from the source C project along with their dependencies, necessitating a prioritization strategy for translation. 
Intuitively, all dependent code units, whether through \textbf{call} or \textbf{refX} relationships, must be translated into Rust before processing any given unit. If circular dependencies exist among code units, these units should be put into one translation unit to be translated together. Therefore, \ourtool{} performs an analysis of the dependency graph before translation to identify translation units and determine their processing order.

\begin{figure}[htb]
    \centering
    \includegraphics[width=0.75\textwidth]{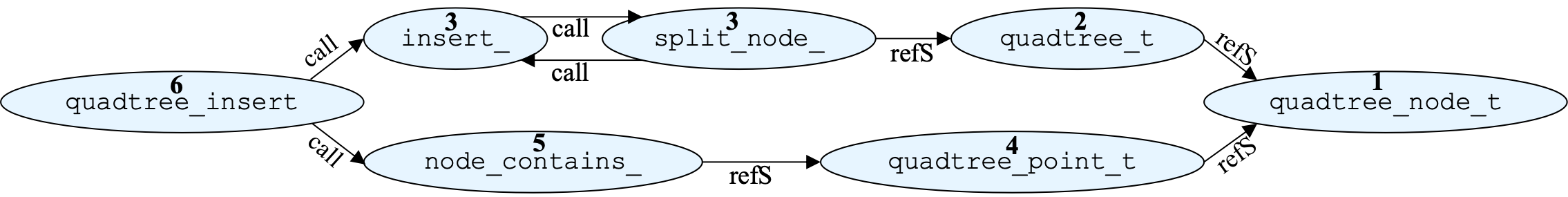}
    \caption{Translation Order Based on Dependency Graph. Numbers Indicate the Translation Order.}
    \label{fig:order}
\end{figure}

To identify translation units, \ourtool{} applies Tarjan's algorithm~\cite{DBLP:journals/siamcomp/Tarjan72} on the code-dependency graph to detect strongly connected components (SCCs). Each SCC, consisting of multiple mutually recursive units or a single independent one, is treated as an individual translation unit. Notably, \ourtool{} removes all manual memory deallocation operations (\eg{} \texttt{free} calls) within functions and eliminates functions solely dedicated to manual deallocation from the translation unit (\eg{} \texttt{quadtree\_point\_free}~\cite{freeFunc}). This addresses the fundamental difference between manual memory management in C and ownership-based automatic reclamation in Rust. Retaining \texttt{free} calls from C code may cause LLMs to generate unsafe Rust. 
However, existing LLM-based translation methods~\cite{DBLP:journals/corr/abs-2412-14234, DBLP:journals/corr/abs-2503-17741, DBLP:journals/corr/abs-2405-11514, DBLP:journals/corr/abs-2409-10506, DBLP:journals/corr/abs-2505-10708, DBLP:journals/ese/HongR25} overlook this critical aspect of manual deallocation.

After partitioning, \ourtool{} condenses each SCC into a single node, converting the code-dependency graph, which may contain cycles, into a directed acyclic graph (DAG). 
A topological sort is subsequently performed on this DAG to establish a bottom-up translation order, ensuring that all dependencies of any given translation unit are translated into Rust before the unit itself.
Fig.~\ref{fig:order} presents a portion of the code-dependency graph extracted from the Quadtree~\cite{quadtree} project, illustrating the translation order derived from this graph (indicated by numerical sequence, starting from \texttt{1}). Functions \texttt{insert\_} and \texttt{split\_node\_} are mutually recursive, thus they are treated as a single translation unit (labeled \texttt{3}) and translated together. The numerical labels represent the final bottom-up translation order, ensuring that each unit is translated only after all its dependencies, including calls and referenced types, have been processed.

\subsubsection{Pointer Semantic Extraction}

For each obtained translation unit, \ourtool{} retrieves pointer semantic knowledge from \kg{}, including pointer-usage information and Rust-oriented annotations. 
This knowledge is incorporated into the ``Pointer Semantic'' section of the translation prompt (Fig.~\ref{fig:trans_prompt}) as \textit{<\texttt{Entity1}, Relation, \texttt{Entity2}>} triples.
For instance, for the parameter \texttt{tree} in the function \texttt{insert\_}, the extracted semantic knowledge is expressed as triples such as \textit{<\texttt{tree}, pointsTo, \texttt{quadtree\_t}>} and \textit{<\texttt{tree}, isA, \texttt{Borrowed}>}.
Furthermore, if two pointer parameters within a function exhibit a \textbf{derivesFrom} relationship and different mutability, \ourtool{} explicitly provides refactoring guidance to LLMs with the following format.

\begin{mdframed}[linecolor=gray,roundcorner=12pt,backgroundcolor=gray!15,linewidth=3pt,innerleftmargin=2pt, leftmargin=0cm,rightmargin=0cm,topline=false,bottomline=false,rightline = false, hidealllines=true]
\textbf{Param\_1} \textit{derivesFrom} \textbf{Param\_2}, where \textbf{Param\_1} requires \textit{Immutable/Mutable} and \textbf{Param\_2} requires \textit{Immutable/Mutable}. Refactor parameters based on actual usage: \textbf{Param\_1} \textit{accessed member}: \textbf{memberName}; \textbf{Param\_2} \textit{accessed member:} \textbf{memberName}.
\end{mdframed}

\subsubsection{Incremental Translation and Validation}
\ourtool{} integrates the translation unit, the pointer semantic knowledge, and the translated Rust context into the translation prompt (see Fig.~\ref{fig:trans_prompt}) to guide LLM in generating the corresponding Rust code. The translated Rust context comprises all dependent code units of the current translation unit that have already been translated into Rust. To control the prompt length, if the Rust code unit is a function, only its signature is included; for other types of code units, the entire code content is incorporated. Furthermore, the translation prompt defines rules (\eg{} prohibiting unsafe code blocks and raw pointers) to guide LLMs in generating safe and idiomatic Rust code. To resolve potential visibility issues caused by cross-module dependencies, the prompt instructs the LLM to annotate all generated Rust entities with the \texttt{pub} modifier.

\begin{figure}[htb]
    \centering
    \includegraphics[width=0.65\textwidth]{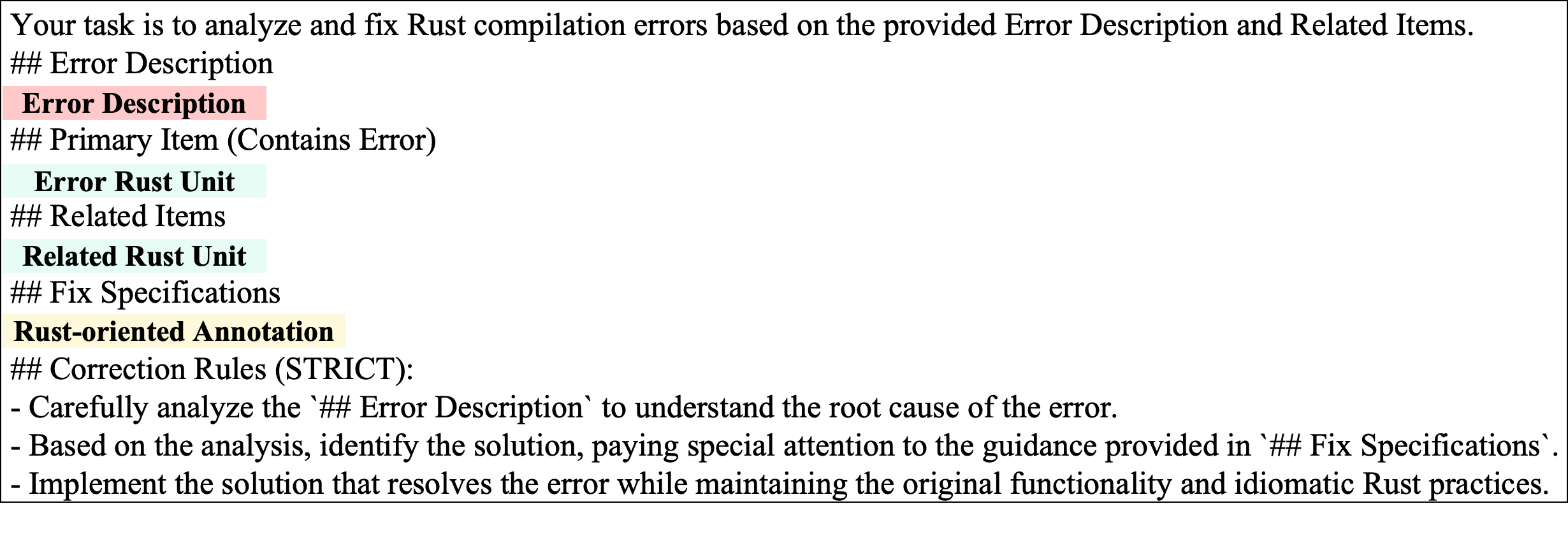}
    \caption{Error Correction Prompt}
    \label{fig:error_prompt}
\end{figure}

\ourtool{} initializes an empty Rust project using the \texttt{cargo new} command at the start of the translation process, providing a container for all translated Rust code units. The initialized project includes only a manifest file (\texttt{Cargo.toml}) and a minimal entry file (\texttt{src/main.rs}).
As each new Rust code unit is generated, \ourtool{} integrates it into the project: units corresponding to C header files (\texttt{.h}) are treated as public modules and saved to \texttt{common/fileName\_mod.rs}, while those from C source files (\texttt{.c}) are placed in \texttt{src/fileName.rs}, with \texttt{fileName} reflecting the original C file name. 
After each integration, \ourtool{} performs compilation verification. Upon encountering errors, it initiates an error correction procedure (see Section~\ref{sec:error_fix}) to prevent error accumulation.  
If no errors occur, \ourtool{} translates the next code unit, repeating until all units are translated.

To preserve and leverage project-level pointer semantics, \ourtool{} dynamically maintains a \textit{Rust copy of the pointer KG} during translation. In this Rust copy, nodes correspond to translated Rust code units (rather than C code units), while edges capture dependency relationships, such as \textbf{call} and \textbf{refX}, inherited from the original C code. Each node is further annotated with metadata, which includes the original C source file information and the pointer semantic knowledge used in translation, to facilitate subsequent integration and error correction.

\subsection{KG-Guided Error Correction}~\label{sec:error_fix}
During error correction, \ourtool{} leverages the Rust copy of our pointer KG 
to provide LLMs with project-level semantic context, rather than relying solely on compiler error messages as in prior methods~\cite{DBLP:journals/corr/abs-2501-14257,DBLP:journals/corr/abs-2405-11514,DBLP:journals/corr/abs-2503-17741,DBLP:journals/corr/abs-2404-18852,DBLP:journals/corr/abs-2412-14234,DBLP:journals/corr/abs-2409-10506,DBLP:journals/ese/HongR25}. The KG assists repair in two main ways: (1) its dependency relations enable precise identification of code units involved in a given error, clarifying inter-component interactions; and (2) each node is enriched with Rust-oriented annotations (\eg{} ownership and mutability), which supply precise program semantics that guide LLMs toward effective fixes and prevent them from getting stuck in unproductive repair loops.

Once a newly translated Rust unit is integrated into the initialized project, \ourtool{} performs compilation verification. If errors are detected, \ourtool{} parses the compiler diagnostics to identify the error code unit and aggregates its errors into a unified description. This description provides the LLM with a holistic view of the problematic unit, facilitating targeted fixes. \ourtool{} then retrieves all units directly linked to the error code unit via \textbf{call} or \textbf{refX} dependencies, along with their Rust-oriented annotations from the Rust copy of our pointer KG. \ourtool{} combines the error code unit, error description, directly related code units, and their Rust-oriented annotations into the correction prompt (see Fig.~\ref{fig:error_prompt}) to instruct LLMs to generate a patch. The patch is integrated back into the Rust project, followed by another compilation verification. This repair cycle repeats until the project compiles successfully or a predefined repair limit is reached.


\begin{figure}[htb]
    \centering
    \includegraphics[width=0.85\textwidth]{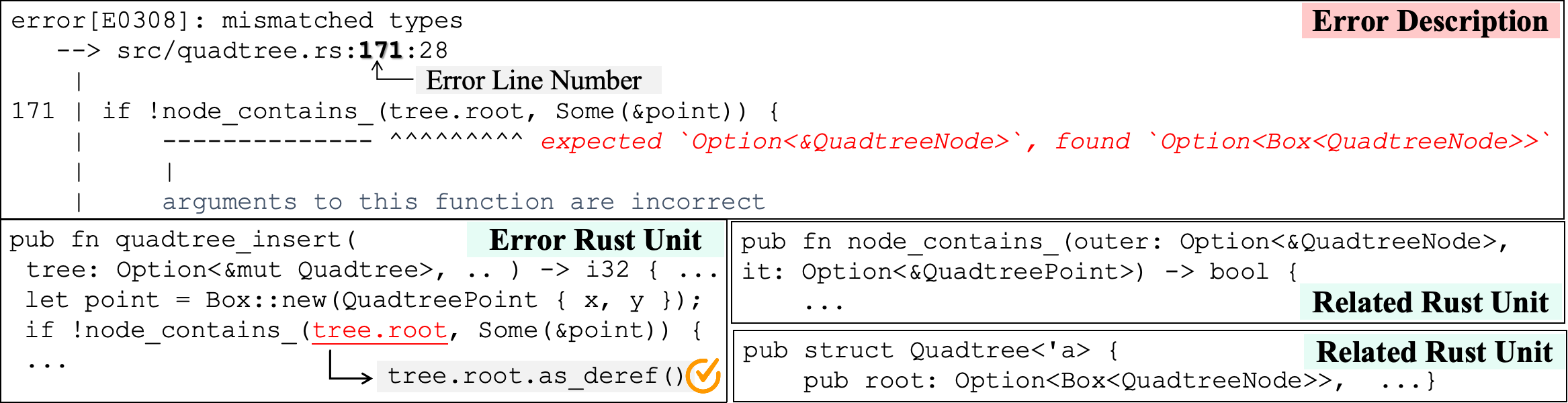}
    \caption{Error Correction Example}
    \label{fig:error_example}
\end{figure}

Fig.\ref{fig:error_example} illustrates a mismatched type error encountered when compiling the Quadtree\cite{quadtree} after integrating \texttt{quadtree\_insert}.
\ourtool{} first pinpoints the error code unit (\texttt{quadtree\_insert}) based on the reported error line, then retrieves directly related units (\eg{} \texttt{node\_contains\_}, \texttt{Quadtree}) along with Rust-oriented annotations (\eg{} <\texttt{root}, \textit{isA}, \textit{Owning}>) from the Rust copy of pointer KG. These elements are incorporated into a correction prompt for the LLM.
The LLM resolves the type mismatch by replacing \texttt{tree.root} with \texttt{tree.root.as\_deref()}, which preserves the ownership and borrowing semantics of the original program design. 
In contrast, relying solely on LLMs for repair based on error descriptions may result in incorrect fixes, such as altering the ownership of the \texttt{root} in \texttt{Quadtree} from \texttt{Option<Box<QuadtreeNode>{}>} to \texttt{Option<\&QuadtreeNode>}, or changing the \texttt{outer} in \texttt{node\_contains\_} from \texttt{Option<\&QuadtreeNode>} to \texttt{Option<Box<QuadtreeNode>{}>}. 
Such incorrect fixes arise because LLMs, lacking a global perspective, tend to focus only on eliminating the current error without preserving the correct ownership and borrowing semantics.

In addition, if compilation errors persist after reaching the maximum repair attempts, \ourtool{} creates a \textit{function stub} to maintain project integrity and compilability. \ourtool{} first reverts the Rust project to the last compilable state before integrating the new Rust code unit, discarding all repair attempts. \ourtool{} then generates a \textit{function stub} as a substitute for the Rust unit that cannot be repaired. 
\zq{Thus, each reversion corresponds to one generated stub.}
Each stub preserves the full function signature (\eg{} ownership, mutability, lifetimes, parameter/return types) and uses \texttt{unimplemented!()} as the body.
\zq{For example, the \textit{function stub} for \texttt{json\_parse\_value} is
\texttt{fn json\_parse\_value(state: Option<\&mut JsonParseState>, is\_global\_object: i32,value: Option<\&mut JsonValue>)\{unimplemented!();\}}}.
This strategy preserves the integrity of dependencies among code units and ensures that the translated project is compilable.

\section{Evaluation}
We evaluate \ourtool{} for project-level C-to-Rust translation with the research questions below:

\begin{itemize}[leftmargin=10pt, topsep=5pt]
\item \textbf{RQ1 (Idiomaticity Evaluation)}: How idiomatic is the Rust code generated by \ourtool{} compared with rule-based and LLM-based rewriting methods?

\item \textbf{RQ2 (Correctness Evaluation)}: How effective is \ourtool{} in translating complete projects, in terms of compilation success and function-level I/O equivalence?

\item \textbf{RQ3 (Ablation Evaluation)}: How do \kg{} knowledge and error-correction phase affect the performance of \ourtool{}?

\item \textbf{RQ4 (Performance Distribution)}: How does \ourtool{} perform across different code complexity (\ie{} lines of code, function dependencies, and pointer counts) and iteration numbers?

\item \zq{\textbf{RQ5 (LLM Generalization Evaluation)}: How does the performance of \ourtool{} vary across different LLM backbones?}

\end{itemize}

\subsection{Experimental Setup}

\subsubsection{Benchmark}

We use real-world C projects from the \crown{} dataset~\cite{crown}, as many projects in this dataset (\eg{} bzip2~\cite{bzip2}, quadtree~\cite{quadtree}) have been widely adopted in prior C-to-Rust translation research~\cite{PR2,DBLP:journals/corr/abs-2501-14257,DBLP:journals/corr/abs-2409-10506,DBLP:journals/pacmpl/EmreBPSDH23}. 
We compile all candidate projects and exclude those with build failures in our environment (\eg{} libcsv~\cite{libcsv}, robotfindskitten~\cite{robotfindskitten}), resulting in a final 16 C projects, as detailed in Table~\ref{table:dataset}.
Moreover, we assess the functional equivalence between each C project and its Rust translation by whether identical test inputs produce identical outputs. We manually construct unit tests for studied projects with the help from automatic generation tools (\eg{} fuzzing and LLMs). 
\zq{Specifically, for each C project, we first use LLMs to generate a test harness for functions with observable behavior. We then apply a fuzzing tool (\eg{} AFL++~\cite{AFLplusplus}) to exercise the harness and collect input–output pairs, from which we select at least three representative test cases per harness. These test cases, together with the harness, are subsequently provided to LLMs (\eg{} ChatGPT~\cite{chatgpt}) to generate the corresponding unit test methods.}
As shown in Table~\ref{table:testCov}, the constructed tests achieve high coverage on the C projects, with 80.0\%–97.7\% line coverage and 87.7\%–100.0\% function coverage, providing a comprehensive functional equivalence evaluation.

\subsubsection{Baselines}

For idiomaticity evaluation (RQ1), we compare \ourtool{} with  \textbf{\crown{}}~\cite{crown} and \textbf{\pr{}}~\cite{PR2}, both of which are representative techniques in rule-based and LLM-based categories. 
\crown{} rewrites the unsafe Rust code generated by \crust{}~\cite{c2rust} based on ownership rules, which has been widely evaluated in prior studies~\cite{DBLP:journals/corr/abs-2404-18852,DBLP:journals/corr/abs-2501-14257,DBLP:journals/corr/abs-2503-12511}; \pr{} uses LLMs to rewrite the Rust code produced by \crust{}, which represents the latest LLM-based method for improving Rust idiomaticity. 

 



\begin{wraptable}{r}{0.5\textwidth}
    \centering
	\small
	\caption{Benchmarks Information. \#DataDecl: total \textit{Struct}, \textit{Enum}, \textit{Union}, \textit{globalVar}, and \textit{Typedef}.}\label{table:dataset}

    \begin{tabular}{l|r|r|r|r}
    \toprule
    Project      & \#File & \#Func. & \#DataDecl & LOC   \\ \midrule
    avl          & 1      & 10     & 1          & 229    \\ \hline
    buffer       & 2      & 23     & 1          & 1207   \\ \hline
    genann       & 6      & 17     & 6          & 2410   \\ \hline
    quadtree     & 5      & 24     & 4          & 1216   \\ \hline
    rgba         & 2      & 13     & 2          & 1855   \\ \hline
    urlparser    & 1      & 21     & 2          & 1379   \\ \hline
    ht           & 1      & 10     & 3          & 264    \\ \hline
    bst          & 1      & 6      & 1          & 154    \\ \hline
    json.h       & 1      & 49     & 16         & 3860   \\ \hline
    libtree      & 1      & 29     & 15         & 2610   \\ \hline
    binn         & 1      & 278    & 13         & 4426   \\ \hline
    bzip2        & 9      & 118    & 48         & 14829  \\ \hline
    heman        & 24     & 334    & 46         & 13762  \\ \hline
    libzahl      & 49     & 59    & 15         & 4655   \\ \hline
    lil          & 2      & 136    & 20         & 5670   \\ \hline
    lodepng      & 1      & 235    & 36         & 14153  \\ \bottomrule
    \end{tabular}
\end{wraptable}


For correctness evaluation (RQ2), we compare our approach with \textbf{\flour{}}~\cite{DBLP:journals/corr/abs-2405-11514}, a representative LLM-based translation approach using fuzzing to improve translation and augmenting LLM-based translation with test cases is a common practice in recent studies~\cite{DBLP:journals/corr/abs-2409-19894, DBLP:journals/pacmse/Yang0YK0LHMJ024, DBLP:journals/corr/abs-2412-08035, DBLP:journals/corr/abs-2412-14234, DBLP:journals/corr/abs-2410-24117}. We exclude other project-level  LLM-based methods from our evaluation, as they are not targeting C-to-Rust translation~\cite{DBLP:journals/corr/abs-2410-24117, guan2025repotransagentmultiagentllmframework, DBLP:journals/corr/abs-2412-08035} or requiring manual intervention during the translation (\eg{} \Syzygy{}~\cite{DBLP:journals/corr/abs-2412-14234}, \RustMap{}~\cite{DBLP:journals/corr/abs-2503-17741}).

For ablation evaluation (RQ3), we compare with four variants(\ie{} \textbf{\ourtoolPA{}}, \textbf{\ourtoolPU{}}, \textbf{\ourtoolRA{}}, and \textbf{\ourtoolError{}}), to analyze the contribution of each component to the correctness of \ourtool{}.
\ourtoolPA{} is a variant of \ourtool{} that excludes the retrieval of pointer semantic knowledge from \kg{} for each translation unit, assessing the impact of pointer-usage information and Rust-oriented annotation on translation quality.
Moreover, \ourtoolPA{} reflects the prevalent bottom-up paradigm for translating C projects into Rust based on the call graph~\cite{DBLP:journals/corr/abs-2412-14234, DBLP:journals/corr/abs-2405-11514, DBLP:journals/corr/abs-2409-10506, DBLP:journals/ese/HongR25}.
\ourtoolPU{} and \ourtoolRA{} are ablated variants that omit pointer-usage information and Rust-oriented annotations, respectively, to assess their necessity during translation.
\ourtoolError{} is identical to \ourtool{} but excludes the error-correction phase, isolating its contribution to the final translation correctness.

\subsubsection{Evaluation Metrics}
We employ two metrics to evaluate idiomaticity: \textbf{Lint Alert Count} and \textbf{Unsafe Usage Count}, both of which have also been used in prior work to evaluate the Rust code idiomaticity~\cite{DBLP:journals/corr/abs-2503-12511,DBLP:journals/corr/abs-2405-11514,DBLP:journals/corr/abs-2404-18852}.
\textbf{Lint Alert Count} is measured using Rust-Clippy~\cite{clippy}, a tool that reports unidiomatic Rust code from multiple perspectives, including style (\eg{} nonstandard naming conventions or unnecessary borrowing), complexity (\eg{} redundant type casts or parentheses), correctness (\eg{} checking if an unsigned integer is greater than 0) and performance (\eg{} cloning values unnecessarily). By collecting the warnings and errors generated by Rust-Clippy on the translated Rust project, we assess its idiomaticity: a lower alert count indicates a more idiomatic translation.
\textbf{Unsafe Usage Count} is obtained by Cargo-geiger~\cite{geiger}, which reports the occurrences of unsafe code within the Rust project. 
Specifically, we record the number of unsafe functions and expressions reported by Cargo-geiger, including \texttt{unsafe fn} declarations, \texttt{unsafe \{\}} blocks, and unsafe operations like raw pointer dereferences from the translated Rust project.
A lower unsafe usage count indicates that the translated code achieves better memory safety guarantees.

We use two metrics to evaluate translation correctness: 
\textbf{Compiled} = $\frac{\textit{compiledFunc}}{\textit{translatedFunc}}$ and 
\textbf{Equiv.} =  $\frac{\textit{equivalentFunc}}{\textit{translatedFunc}}$.  
Here, \textit{translatedFunc} denotes the total number of source C functions translated into Rust, excluding functions solely dedicated to memory deallocation (\eg{} \texttt{quadtree\_point\_free}~\cite{freeFunc} in Quadtree). 
\zq{\textit{compiledFunc} refers to translated functions whose bodies are not replaced by the \texttt{unimplemented!()} macro (\ie{} functions stubbed after exhausting repair attempts are excluded); therefore, \textbf{Compiled} reaches 100\% if and only if no function stub is generated.}
\textit{equivalentFunc} represents the number of translated functions that successfully pass their corresponding unit tests.
It is worth noting that the \textbf{Equiv.} depends on test coverage. Functions not covered by unit tests are automatically considered non-equivalent.

\begin{wraptable}{r}{0.45\textwidth}
    \centering
	\small
    \vspace{-3mm}
	\caption{\zq{Unit Test  Coverage}}\label{table:testCov}
\begin{tabular}{l|r|r}
\toprule
Project      & Line Cov.(\%) & Func. Cov.(\%) \\ \midrule
avl          & 81.4              & 100.0              \\ \hline
buffer       & 91.7              & 92.9                \\ \hline
genann       & 92.4              & 95.5                 \\ \hline
quadtree     & 94.0                & 100.0                \\ \hline
rgba         & 92.5              & 100.0            \\ \hline   
urlparser        & 97.7              & 100.0           \\ \hline
ht               & 93.6                 & 100.0         \\ \hline
bst & 96.2                 & 100.0         \\ \hline
json.h           & 82.7              & 95.8         \\ \hline
libtree          & 80.0            & 100.0  \\ \hline\hline
\zq{binn}      & \zq{89.2} & \zq{94.4}  \\ \hline
\zq{bzip2}     & \zq{80.6} & \zq{87.7}  \\ \hline
\zq{heman}     & \zq{91.2} & \zq{99.0}  \\ \hline
\zq{libzahl}   & \zq{80.2} & \zq{100.0} \\ \hline
\zq{lil}       & \zq{80.2} & \zq{100.0} \\ \hline
\zq{lodepng}   & \zq{80.3} & \zq{100.0} 

\\ \bottomrule
\end{tabular}

\end{wraptable}

\subsubsection{Implementation}

\textbf{\ourtool{} Implementation}.
For \kg{} construction, we first use \texttt{Clang -E} (version \textit{14.0.6}) to perform macro expansion on the given C project. 
Subsequently, we perform static analysis with \textit{Doxygen}~\cite{doxygen}, a multi-language tool that extracts structured information from code bases, to obtain code units and dependencies, constructing the code-dependency graph.
To extract pointer-usage information and Rust-oriented annotations, we first compile the C project into LLVM Intermediate Representation (IR) using the \texttt{clang -emit-llvm} command. We then employ \textit{SVF}~\cite{svf} (version 2.9), a static value-flow analysis framework for C/C++, to perform a context-sensitive and field-sensitive analysis of the IR.
For the KG-guided translation phase, the code-dependency graph determines translation units and their order, while the pointer semantic knowledge retrieved from the KG provides contextual guidance for LLMs to translate each unit.
Translated Rust code is immediately integrated into a Rust project initialized via \texttt{cargo new} and verified using \texttt{cargo check} (version \textit{1.90.0-nightly}). If compilation errors are detected, \ourtool{} parses the diagnostic messages and supplies them to LLMs, which iteratively repair the code. The iterative repair process is limited to five cycles, in accordance with empirical findings that demonstrate this threshold maximizes correction efficiency~\cite{DBLP:journals/corr/abs-2409-19894, DBLP:journals/pacmse/Yang0YK0LHMJ024,DBLP:journals/pacmse/Yuan0DW00L24}.

\textbf{Baseline Implementation.} 
We evaluate against \crown{}~\cite{crown}, \pr{}~\cite{PR2}, and \flour{}~\cite{DBLP:journals/corr/abs-2405-11514} using their released implementations.
Since \flour{} partitions C projects into translation units based on dependencies (up to depth 4) but does not define their translation order, we determine it bottom-up from the call dependency graph and use Restart, the best-performing strategy in their experiments, for repair. Additionally, we reconfigure \pr{} and \flour{} to utilize the same LLM as \ourtool{} for comparability.

\textbf{LLMs.}
We use ChatGPT-4o~\cite{chatgpt} as the default backbone LLM, as it has been widely adopted for a variety of code-related tasks~\cite{DBLP:journals/access/AlmanasraS25, DBLP:journals/corr/abs-2502-07399, 11061239, DBLP:journals/corr/abs-2412-17744}.
\zq{Moreover, \claude{}~\cite{Claude} and \gemini{}~\cite{Gemini} are used as alternative backbone LLMs to evaluate the generalization performance of \ourtool{} across different models.}
To ensure experimental consistency and reproducibility, we set the model parameters to ``temperature=0'' to control randomness of results.

\subsubsection{Experimental Procedure}
The evaluation methodology for each research question is as follows.

\textbf{RQ1 (Idiomaticity Evaluation).}
We compare the translation idiomaticity of \ourtool{} to \crown{} and \pr{} with the Lint Alert Count and Unsafe Usage Count of their translated Rust projects.

\textbf{RQ2 (Correctness Evaluation).}
We compare the translation correctness of \ourtool{} to \flour{} with \%Compiled and \%Equiv. on their translated Rust projects.

\textbf{RQ3 (Ablation Evaluation).}
We evaluate the translation correctness of \ourtool{} to four variants of \ourtool{} (\ie{} \ourtoolPA{}, \ourtoolPU{}, \ourtoolRA{}, \ourtoolError{}) with the \%Compiled and \%Equiv. on  their translated Rust projects. 

\textbf{RQ4 (Performance Distribution Analysis).}
For each source C function, we analyze its code complexity, including lines of code (LOC),  the number of dependent code units (both direct and indirect), and the number of pointers. We then analyze how \ourtool{} and the baseline \flour{} perform across function groups of different code complexity. 
Additionally, we analyze the performance of \ourtool{} in different error-correction iterations.

\zq{\textbf{RQ5 (LLM Generalization Evaluation).}}
\zq{To evaluate the generalization capability of \ourtool{} across different LLM backbones, we replace the default ChatGPT-4o with two alternative LLMs, \ie{} \claude{}~\cite{Claude} and \gemini{}~\cite{Gemini}. For each LLM setting, we compare the effectiveness of two variants: (i) \ourtoolPA{} (without pointer semantics) and (ii) \ourtool{} (full framework), to assess the impact of LLM choice on translation performance.}

\subsection{Experimental Results}

\subsubsection{RQ1: Idiomaticity Evaluation}

Table~\ref{table:rq1} reports the distribution of \textit{Lint Alert Count} and \textit{Unsafe Usage Count} across 16 Rust projects translated by \crown{}, \pr{}, and \ourtool{}. Lower values indicate more idiomatic and safer Rust code. Overall, \ourtool{} produces more idiomatic and safer translations than \crown{} and \pr{}. Across the 16 projects, \ourtool{} reduces the Lint Alert Count by 94.9\% compared to \crown{} (from 6,802 to 349) and by 91.6\% compared to \pr{} (from 4,135 to 349). Similarly, \ourtool{} reduces the Unsafe Usage Count by 99.9\% against both \crown{} (from 141,866 to 85) and \pr{} (from 134,185 to 85). Wilcoxon signed-rank tests~\cite{rosner2006wilcoxon} show that \ourtool{} significantly lowers both metrics compared with \crown{} and \pr{} ($p < 0.001$).

\begin{table*}[htb]
    \centering
	\small
	\caption{Evaluation of Rust Code Idiomaticity: \ourtool{} vs. \crown{}, \pr{}}\label{table:rq1}
    \setlength{\tabcolsep}{2pt}
    \begin{adjustbox}{width=1.0\columnwidth}
    
\begin{tabular}{c|cccccccccccc|ccc}
\toprule
\multirow{3}{*}{Project} & \multicolumn{12}{c|}{Lint Alert Count}                                                                                                                     & \multicolumn{3}{c}{Unsafe Usage Count}                                     \\ \cline{2-16} 
                             & \multicolumn{3}{c|}{Style Lints} & \multicolumn{3}{c|}{Complexity Lints} & \multicolumn{3}{c|}{Correctness Lints} & \multicolumn{3}{c|}{Performance Lints} & \multirow{2}{*}{\crown{}} & \multirow{2}{*}{\pr{}} & \multirow{2}{*}{\ourtool{}} \\ \cline{2-13}
                             & \crown{}    & \pr{}     & \ourtool{}   & \crown{}      & \pr{}      & \ourtool{}     & \crown{}      & \pr{}      & \ourtool{}      & \crown{}      & \pr{}      & \ourtool{}      &                        &                      &                            \\ \midrule
avl                          & 17       & 30      & 3           & 52         & 0        & 0             & 0          & 0        & 0              & 1          & 0        & 0              & 497                    & 309                  & 0                          \\
buffer                       & 54       & 62      & 4           & 73         & 6        & 0             & 0          & 0        & 0              & 2          & 0        & 0              & 1678                   & 1531                 & 0                          \\
genann                       & 44       & 68      & 3           & 32         & 6        & 2             & 0          & 0        & 0              & 0          & 0        & 0              & 3851                   & 3822                 & 0                          \\
quadtree                     & 60       & 56      & 0           & 146        & 7        & 4             & 0          & 0        & 0              & 4          & 0        & 0              & 1989                   & 1748                 & 0                          \\
rgba                         & 168      & 175     & 6           & 4          & 3        & 0             & 0          & 0        & 0              & 0          & 0        & 0              & 1302                   & 1257                 & 0                          \\
urlparser                    & 64       & 66      & 7           & 73         & 1        & 0             & 0          & 0        & 0              & 0          & 0        & 1              & 1725                   & 1719                 & 0                          \\
ht                           & 23       & 35      & 2           & 44         & 0        & 1             & 0          & 0        & 0              & 1          & 0        & 0              & 468                    & 417                  & 0                          \\
bst                          & 11       & 20      & 1           & 31         & 0        & 1             & 0          & 0        & 0              & 1          & 0        & 0              & 239                    & 168                  & 0                          \\
json.h                       & 215      & 344     & 31          & 472        & 10       & 2             & 2          & 2        & 1              & 0          & 0        & 0              & 9101                   & 8049                 & 0                          \\
libtree                      & 44       & 49      & 2           & 173        & 9        & 2             & 0          & 0        & 0              & 0          & 0        & 0              & 4429                   & 4121                 & 0                          \\ \hline
binn                         & 445      & 560     & 26          & 269        & 11       & 11            & 0          & 0        & 0              & 0          & 0        & 1              & 6489                   & 6089                 & 35                         \\
bzip2                        & 220      & 230     & 57          & 380        & 128      & 37            & 47         & 47       & 3              & 0          & 0        & 10             & 31394                  & 30751                & 20                         \\
heman                        & 536      & 799     & 17          & 683        & 18       & 5             & 3          & 3        & 2              & 0          & 0        & 0              & 31894                  & 30139                & 14                         \\
libzahl                      & 160      & 202     & 20          & 113        & 31       & 8             & 2          & 2        & 0              & 0          & 0        & 0              & 6436                   & 6265                 & 0                          \\
lil                          & 273      & 415     & 24          & 869        & 3        & 4             & 2          & 2        & 0              & 0          & 0        & 0              & 12858                  & 11183                & 0                          \\
lodepng                      & 456      & 697     & 30          & 531        & 36       & 17            & 2          & 2        & 4              & 0          & 0        & 0              & 27516                  & 26617                & 16                         \\ \midrule
Total                        & 2790     & 3808    & 233         & 3945       & 269      & 94            & 58         & 58       & 10             & 9          & 0        & 12             & 141866                 & 134185               & 85                         \\ \bottomrule
\end{tabular}
    \end{adjustbox}
\end{table*}

Regarding Lint Alert Count, \ourtool{} consistently produces fewer alerts across 16 translated Rust projects than \crown{} and \pr{}. 
For instance, in \textit{rgba}, \ourtool{} reduces the lint by 96.5\% vs. \crown{} (from 172 to 6) and by 96.6\% vs. \pr{} (from 178 to 6). 
Further analysis shows that \textit{rgba} code translated by \ourtool{} triggers only six style-related lints, half of which stem from using ``\textit{s.chars().nth(0)}''. Clippy recommends replacing this with \texttt{.next()}, which avoids unnecessary index checks.
In contrast, \pr{} produces 178 style-related lints, 149 of which arise from unnecessary \texttt{let} bindings and redundant \texttt{return} statements, patterns that are considered unidiomatic in Rust~\cite{DBLP:journals/pacmpl/HongR24}.
For complexity-related lints, \ourtool{} reports 1–3 more alerts than \pr{} in four projects (\textit{ht}, \textit{bst}, \textit{bzip2}, \textit{lil}). 
For example, in \textit{ht}, \ourtool{} triggers a lint by using a \texttt{loop} + \texttt{if let} pattern, whereas Clippy recommends \texttt{while let} for clarity. By contrast, \pr{} employs a \texttt{while} loop by directly mapping the syntax of the source C code and thus avoids the lint; however, the loop body contains extensive unsafe code usage.
For performance-related lints, \ourtool{} reports 10 alerts in the \textit{bzip2} project.
Further analysis shows that 9 alerts are Clippy warnings for ``unnecessary use of \texttt{to\_string()}''. In fact, the \texttt{to\_string()} calls are used to convert the return value of \texttt{Cow<str>}, whether a borrowed \texttt{\&str} or an owned \texttt{String}, into an owned \texttt{String}, ensuring type safety.

Regarding Unsafe Usage Count, \ourtool{} produces Rust with zero unsafe usages for projects under 4k LOC (\eg{} \textit{avl}). 
In the remaining projects (\eg{} binn), the amount of unsafe code is reduced by 99.9\% both against both \crown{} and \pr{}, showing that \ourtool{} produces more safer Rust code. Further analysis reveals that \crown{} and \pr{} often use \texttt{pub unsafe functions}, \texttt{unsafe \{\}} blocks, and raw pointer dereferencing, all of which increase memory safety risks.  The unsafe code in \ourtool{} mainly occurs in low-level memory operations translation (\eg{} null-pointer checks). 
For example, in \textit{bzip2}, C code verifies whether the pointer returned by \texttt{malloc} is \texttt{null} before proceeding. \ourtool{} translates \texttt{malloc} into Rust’s allocation function (\texttt{alloc}) and replicates the \texttt{null} check, which requires an \texttt{unsafe} block.

\begin{wraptable}{r}{0.5\textwidth}
 \centering
 \small
\caption{Correctness Evaluation}\label{table:correct}
\resizebox{\linewidth}{!}{
\begin{tabular}{c|cc|cc}
\toprule
\multirow{2}{*}{Project} & \multicolumn{2}{c|}{\flour{}}              & \multicolumn{2}{c}{\ourtool{}}             \\ \cline{2-5} 
                         & \%Compiled & \%Equiv. & \%Compiled & \%Equiv. \\ \hline
avl                      & 90.0     & 70.0   & 100.0    & 100.0  \\ \hline
buffer                   & 100.0    & 70.0   & 100.0    & 75.0   \\ \hline
genann                   & 37.5     & 11.1   & 100.0    & 77.8   \\ \hline
quadtree                 & 70.6     & 43.8   & 100.0    & 87.5   \\ \hline
rgba                     & 100.0    & 100.0  & 100.0    & 100.0  \\ \hline
urlparser                & 78.9     & 10.5   & 100.0    & 36.8   \\ \hline
ht                       & 100.0    & 100.0  & 100.0    & 100.0  \\ \hline
bst                      & 100.0    & 100.0  & 100.0    & 100.0  \\ \hline
json.h                   & 10.4     & 10.4   & 93.8     & 68.8   \\ \hline
libtree                  & 11.1     & 7.4    & 88.9     & 70.4   \\ \hline 
\textbf{Ave.}                     & 69.9      & 52.3    & 98.3      & 81.6   \\ \hline \hline 
\zq{binn}                     & \zq{100.0}   & \zq{5.8}   & \zq{92.4}    & \zq{72.3}   \\ \hline
\zq{bzip2}                    & \zq{65.6}    & \zq{8.5}   & \zq{88.0}     & \zq{62.7}   \\ \hline
\zq{heman}                    & \zq{54.5}    & \zq{20.1}  & \zq{93.4}    & \zq{54.5}   \\ \hline
\zq{libzahl}                  & \zq{39.0}    & \zq{30.5}  & \zq{69.5}    & \zq{69.5}  \\ \hline
\zq{lil}                      & \zq{68.3}    & \zq{10.8}  & \zq{80.1}    & \zq{80.0}   \\ \hline
\zq{lodepng}                  & \zq{56.6}    & \zq{9.2}   & \zq{91.9}    & \zq{68.4}  \\ \hline
\textbf{\zq{Ave.}}                 & \zq{64.0}    & \zq{14.2}    & \zq{85.9}      & 67.9
\\ \bottomrule
\end{tabular}
}
\end{wraptable}

\subsubsection{RQ2: Correctness Evaluation}
As shown in Table~\ref{table:correct}, \ourtool{} substantially outperforms \flour{} across all 16 evaluated projects. Overall, \ourtool{} substantially outperforms \flour{}. Particularly, the \textit{avl}, \textit{rgba}, \textit{ht}, and \textit{bst} projects translated by \ourtool{} pass all test methods.

For the projects with LoC < 4K, \ourtool{} outperforms \flour{}, achieving average improvements of 28.4\% (= 98.3\% - 69.9\%) in compilation success rate and 29.3\% (= 81.6\% - 52.3\%) in functional equivalence.
This performance advantage can be primarily attributed to the project-specific C–Rust pointer KG constructed by \ourtool{}. The KG decomposes each project into independently translatable units, thereby alleviating the performance degradation that \flour{} suffers from as the context length increases in monolithic translation. Moreover, the KG enriches each translation unit with explicit pointer semantic information (\eg{} Ownership), which guides LLMs toward more accurate translations.
For the \textbf{\textit{urlparser}} project, both \ourtool{} and \flour{} achieve relatively low functional equivalence.
The reason is that the critical function \texttt{get\_parser} (which is invoked by other 10 functions) is not correctly translated. After manually fixing \texttt{get\_parser}, the equivalence rises from 36.8\% to 94.7\% for \ourtool{}, and from 10.5\% to 36.8\% for \flour{}.

For the larger projects ($4\text{K} < \text{LoC} < 15\text{K}$), \ourtool{} remains effective and substantially outperforms \flour{}, improving the average compilation success rate from 64.0\% to 85.9\% and functional equivalence from 14.2\% to 67.9\%.
As project size increases, pointer semantics become dispersed across multiple functions and files, making it difficult for function-local translation approaches to infer correct Rust semantics.  \ourtool{} mitigates this issue by building a pointer KG before translation to inject globally consistent pointer semantics into each translation unit.
For example, in \texttt{bzip2}, the \texttt{EState} struct contains pointer fields (\texttt{block}, \texttt{mtfv}, \texttt{ptr}) that are derived from the base arrays \texttt{arr1} and \texttt{arr2} at fixed offsets during initialization in \texttt{BZ2\_bzCompressInit}. Without capturing this derivation, \flour{} initializes these fields as null pointers and omits the required initialization logic; subsequent accesses in functions such as \texttt{generate\_mtf\_values} then operate on uninitialized pointers and produce incorrect output. In contrast, \ourtool{}’s pointer KG captures the derivation between \texttt{block} and \texttt{arr2} during pre-translation analysis, enabling correct initialization and preserving the functional correctness of the compression pipeline.

\subsubsection{RQ3: Ablation Evaluation}

Table~\ref{table:ablation} shows the contribution of knowledge in \kg{} and error correction to \ourtool{} translation performance.

Both pointer usage information and Rust-oriented annotations from \kg{} are essential to improving the translation performance of \ourtool{}. Incorporating global-perspective pointer semantics from the KG for each translation unit raises the average equivalence rate from 59.5\% (\ourtoolPA{}, without pointer semantics) to 81.6\% (\ourtool{}), a gain of 22.1\%. 
This improvement occurs because, without global pointer semantics, LLMs cannot capture cross-unit pointer usage, and error corrections proceed without factual guidance. For example, in the \textit{avl}, after \ourtoolPA{} translates \texttt{deleteNode}, a type mismatch arises when calling \texttt{height}: ``\texttt{Option<\&Node>} is expected, but \texttt{Option<\&Box<Node>{}>} is found''. Without factual knowledge of pointer ownership, \ourtoolPA{} makes incorrect modifications, such as changing the \texttt{height} parameter type to \texttt{Option<\&Box<Node>{}>}.
While this resolves the mismatch between \texttt{deleteNode} and \texttt{height}, it triggers type errors in other modules that call \texttt{height}.
Consequently, \ourtoolRA{} gets stuck in ineffective repair loops and fail the translation. When the ownership types of the parameters for \texttt{height} and \texttt{deleteNode} are known, LLMs correctly apply \texttt{.as\_deref()} to resolve the error.

\begin{table*}[htb]
    \centering
	\small
	\caption{Ablation Evaluation : \ourtool{} vs. \ourtool{} Variants}\label{table:ablation}
 
    \begin{adjustbox}{width=1.0\columnwidth}
    
\begin{tabular}{ccccccccccccc}
\toprule
\multicolumn{2}{c}{}                                       & avl            & buffer        & genann        & quadtree      & rgba           & urlparser     & ht             & bst            & json.h        & libtree       & Ave.          \\ \midrule

\multirow{2}{*}{\ourtoolPA}    & \%Compiled & 90.0           & 95.5          & 87.5          & 88.2          & 100.0          & 94.7          & 88.9           & 80.0           & 79.2          & 88.9          & 89.3          \\
                                              & \%Equiv.   & 70.0           & 45.0          & 66.7          & 56.2          & 100.0          & 15.8          & 55.5           & 80.0           & 54.2          & 51.9          & 59.5          \\ \hline
\multirow{2}{*}{\ourtoolPU}    & \%Compiled & 80.0           & 95.5          & 81.2          & 94.1          & 100.0          & 100.0         & 77.8           & 80.0           & 52.1          & 85.2          & 84.6          \\
                                              & \%Equiv.   & 40.0           & 20.0          & 66.7          & 50.0          & 91.7           & 26.3          & 55.6           & 80.0           & 33.3          & 59.3          & 52.9          \\ \hline
\multirow{2}{*}{\ourtoolRA}    & \%Compiled & 100.0          & 95.5          & 93.8          & 70.6          & 100.0          & 100.0         & 88.9           & 100.0          & 56.3          & 74.1          & 87.9          \\
                                              & \%Equiv.   & 100.0          & 25.0          & 66.7          & 56.3          & 100.0          & 21.1          & 66.7           & 100.0          & 35.4          & 48.1          & 61.9          \\ \hline
\multirow{2}{*}{\ourtoolError} & \%Compiled & 80.0           & 86.4          & 75.0          & 70.6          & 92.3           & 52.6          & 33.3           & 80.0           & 41.7          & 48.1          & 66.0          \\
                                              & \%Equiv.   & 40.0           & 60.0          & 66.7          & 56.2          & 91.7           & 15.8          & 33.3           & 80.0           & 31.3          & 33.3          & 50.8          \\ \hline \hline
\multirow{2}{*}{\ourtool}      & \%Compiled & 100.0          & 100.0         & 100.0         & 100.0         & 100.0          & 100.0         & 100.0          & 100.0          & 93.8          & 88.9          & 98.3          \\
                                              & \%Equiv.   & \textbf{100.0} & \textbf{75.0} & \textbf{77.8} & \textbf{87.5} & \textbf{100.0} & \textbf{36.8} & \textbf{100.0} & \textbf{100.0} & \textbf{68.8} & \textbf{70.4} & \textbf{81.6} \\ \bottomrule
\end{tabular}

    \end{adjustbox}
\end{table*}

Providing only pointer-usage information (\ourtoolRA{}) or only Rust-oriented annotations (\ourtoolPU{}) fails to match the translation performance of \ourtool{}.
For instance, in \textit{buffer}, the \%Equiv. for \ourtoolPU{} and \ourtoolRA{} is 55\% and 50\% lower than \ourtool{}, respectively.
Take the member \texttt{char *data} in \texttt{buffer\_t} struct as an example. With only pointer-usage information, LLM fails to infer ownership and translates the field as \texttt{data: Option<*mut u8>}, causing a mismatched-types error at the use site of \texttt{buffer\_length}. With only Rust-oriented annotations, LLM misses usage patterns (\eg{} \texttt{self->data=calloc(n+1,1)}) and translates \texttt{char *data} as \texttt{data: Option<Box<u8>{}>} instead of \texttt{data: Option<Box<[u8]>{}>}, losing the variable-length semantics. Introducing the error correction mechanism raises the average equivalence rate from 50.8\% (\ourtoolError{}) to 81.6\% (\ourtool{}), a 30.8\% gain, demonstrating that KG-guided error correction is effective to improve the translation performance of \ourtool{}.

\subsubsection{RQ4: Performance Distribution Analysis}

Table~\ref{table:errorAna} shows the \%Equiv. distribution of \ourtool{} and the baseline \flour{} across different code complexity (\ie{} grouped by LOC, dependencies, and pointer counts). While \flour{} performs much more poorly (\eg{}  equivalence rates sharply decline to zero) as code complexity increases, \ourtool{} overall exhibits a stable  performance across different code complexity.  For example, as pointer counts rise from [15, 19] to [20, 50),  equivalence rates for \ourtool{} increase from 42.9\% to 57.1\%, while \flour{} drops from 14.3\% to 0\%; similarly, when the number of dependent units per function increases from [12, 15] to [16, 44), equivalence rates for \ourtool{} increase from 50.0\% to 66.7\%, but \flour{} remains at zero. In summary, the results indicate the potential and stable effectiveness of \ourtool{} in translating large-scale projects and complex function calls, which benefits from our \kg{}. 

\begin{table*}[htb]
    \centering
	\small
	\caption{Performance Factor Differences: \ourtool{} vs. \flour{} (\%Equiv. Analysis)}\label{table:errorAna}
    \setlength{\tabcolsep}{3pt}
    \begin{adjustbox}{width=1.0\columnwidth}

\begin{tabular}{ccc|ccc|ccc|cc}
\toprule
\multicolumn{3}{c|}{LOC}                                                                 & \multicolumn{3}{c|}{Dependency Count}                                                    & \multicolumn{3}{c|}{Pointer Count}                                                       & \multirow{2}{*}{\#Iterations} & \multirow{2}{*}{\ourtool{} (\%)} \\ \cline{1-9}
Groups               & \ourtool{} (\%) & \flour{} (\%) & Groups               & \ourtool{} (\%) & \flour{} (\%) & Groups               & \ourtool{} (\%) & \flour{} (\%) &                              &                                                   \\ \midrule
(0, 14{]}        & 84.3                             & 50.0                           & {[}0, 3{]}   & 87.5                             & 53.6                           & {[}0, 4{]}   & 92.2                             & 54.4                           & [0, 1]                         & 84.0                                              \\
{[}15, 29{]} & 61.1                             & 44.4                           & {[}4, 7{]}   & 56.2                             & 12.5                           & {[}5, 9{]}   & 68.6                             & 31.4                           & 2                            & 61.5                                              \\
{[}30, 44{]} & 77.8                             & 22.2                           & {[}8, 11{]}  & 33.3                             & 0.0                            & {[}10, 14{]} & 50.0                             & 18.8                           & 3                            & 40.0                                              \\
{[}45, 59{]} & 60.0                             & 0.0                            & {[}12, 15{]} & 50.0                             & 0.0                            & {[}15, 19{]} & 42.9                             & 14.3                           & 4                            & 71.4                                              \\
{[}60, 390)      & 53.3                             & 0.0                            & {[}16, 44)       & 66.7                             & 0.0                            & {[}20, 50)       & 57.1                             & 0.0                            & 5                            & 0.0                                               \\ \bottomrule
\end{tabular}

    \end{adjustbox}
\end{table*}

Furthermore, the column ``\#Iterations'' in Table~\ref{table:errorAna} further shows the \%Equiv. distribution of \ourtool{} across different correct iterations. Particularly, we find that the correction beyond four iterations (\ie{} five-iteration correction) would not bring additional performance increase. It shows that \ourtool{} can effectively fix the wrong translation within a fixed number of iterations, confirming that our current setting (\ie{} limit the maximum iteration number to 5) is sufficient.

\subsubsection{\zq{RQ5: LLM Generalization Evaluation}}

Table~\ref{table:generalization} presents the effectiveness of \ourtool{} across different backbone LLMs (\ie{} \claude{} and \gemini{}). Overall, \ourtool{} consistently improves performance under all evaluated LLMs. Compared with \ourtoolPA{} (without pointer semantics), \ourtool{} achieves average gains of 1.7\% and 2.7\% in compilation rate, and 23.3\% and 25.3\% in functional equivalence when powered by \claude{} and \gemini{}, respectively. These improvements indicate that the pointer KG provides a model-agnostic enhancement by supplying explicit ownership and mutability semantics for pointer-typed fields and parameters, enabling LLMs to generate idiomatic Rust types (\eg{} \texttt{Option<Box<[u8]>{}>} instead of \texttt{*mut u8}). 
Without such semantic guidance, LLMs conservatively translate pointers as raw pointers, which often results in failures in memory management functions (\eg{} \texttt{buffer\_resize}) and propagates to downstream functional errors. The improvement is most pronounced for projects like \texttt{urlparser} (from 15.8\% to 89.5\%/94.7\%) and \texttt{buffer} (from 35.0\%/45.0\% to 75.0\%/90.0\%), where correct translation of struct fields with implicit ownership semantics is critical for functional correctness.

\begin{table*}[htb]
    \centering
	\small
	\caption{Generalization of \ourtool{}}\label{table:generalization}
 
    \begin{adjustbox}{width=1.0\columnwidth}
\begin{tabular}{ccc|c|c|c|c|c|c|c|c|c|c|c}
\toprule
\multicolumn{3}{c|}{}                                                                                             & avl     & buffer  & genann & quadtree & rgba    & urlparser & ht      & bst     & json.h  & libtree & Ave. \\ \hline
\multicolumn{1}{c|}{\multirow{4}{*}{Claude}} & \multicolumn{1}{c|}{\multirow{2}{*}{\ourtoolPA{}}} & \%Compiled & 100.0 & 95.5  & 81.2 & 94.1   & 100.0 & 100.0   & 100.0 & 100.0 & 89.6  & 100.0 & 96.0 \\ 
\multicolumn{1}{c|}{}                             & \multicolumn{1}{c|}{}                            & \%Equiv.   & 100.0 & 35.0  & 33.3 & 87.5   & 91.7  & 15.8    & 88.9  & 80.0  & 50.0  & 40.7 & 62.3 \\ \cline{2-14} 
\multicolumn{1}{c|}{}                             & \multicolumn{1}{c|}{\multirow{2}{*}{\ourtool{}}}   & \%Compiled & 100.0 & 100.0 & 87.5 & 100.0  & 100.0 & 100.0   & 100.0 & 100.0 & 89.6  & 100.0 & 97.7\\  
\multicolumn{1}{c|}{}                             & \multicolumn{1}{c|}{}                            & \%Equiv.   & 100.0 & 75.0  & 77.8 & 87.5   & 100.0 & 89.5    & 88.9  & 100.0 & 75.0  & 62.3  & 85.6 \\ \midrule
\multicolumn{1}{c|}{\multirow{4}{*}{Gemini}} & \multicolumn{1}{c|}{\multirow{2}{*}{\ourtoolPA{}}} & \%Compiled & 100.0 & 100.0 & 87.5 & 100.0  & 92.3  & 94.7    & 88.9  & 100.0 & 100.0 & 74.1 & 93.8 \\  
\multicolumn{1}{c|}{}                             & \multicolumn{1}{c|}{}                            & \%Equiv.   & 100.0 & 45.0  & 55.6 & 87.5   & 91.7  & 15.8    & 44.4  & 80.0  & 77.1  & 40.7  & 63.8\\ \cline{2-14} 
\multicolumn{1}{c|}{}                             & \multicolumn{1}{c|}{\multirow{2}{*}{\ourtool{}}}   & \%Compiled & 100.0 & 100.0 & 87.5 & 100.0  & 100.0 & 100.0   & 100.0 & 100.0 & 100.0 & 77.8  &  96.5\\ 
\multicolumn{1}{c|}{}                             & \multicolumn{1}{c|}{}                            & \%Equiv.   & 100.0 & 90.0  & 77.8 & 87.5   & 100.0 & 94.7    & 88.9  & 100.0 & 81.3  & 70.4 &  89.1
\\ \bottomrule
\end{tabular}
    \end{adjustbox}
    
\end{table*}
\section{Discussion}
\subsection{Case Study} 
We further manually investigate the translation errors of \ourtool{} and check how many manual effort are required to fix them. In particular, we manually fix incorrectly-translated functions in projects \textit{urlparser} and \textit{libtree} and measure the fixing effort with the two metrics, including  (i) \textit{DSR@k} (Debugging Success Rate)~\cite{DBLP:conf/emnlp/YanTLCW23}, which measures whether a translated function can be fixed to pass the tests within k repair attempts, with each test run counting as one attempt (\ie{} k = 1 in DSR@K); and (ii) \textit{CodeBLEU}~\cite{DBLP:journals/corr/abs-2009-10297}, which evaluates the similarity between the repaired Rust function and translation outputs of \ourtool{}.

Based on the case analysis, we find that functions with translation errors generally require minimal manual effort. Most manually-repaired functions share high similarity with \ourtool{}’s translation, with average CodeBLEU scores of 0.88 for both \textit{urlparser} and \textit{libtree}, indicating that fixing requires only small-scale adjustments. Moreover, the average DSR@k is 1.0 for \textit{urlparser} and 1.3 for \textit{libtree}, showing that most functions can be correctly repaired in one or at most two attempts. In addition, we also find that not all test-failing functions require fixes, as some failures result from error propagation: the tested functions may call other mistranslated functions and fail indirectly. For example, in \textit{urlparser}, 12 unit tests fail, but only 2 functions require actual repair. Specifically, the function \textit{get\_part} is invoked by other 10 functions that all fail on tests; once \textit{get\_part} is repaired, these 10 functions pass without further modifications.  
Therefore, we believe that even for the bad cases, the translation generated by \ourtool{} can still serve as valuable references for developers, which can be fixed into correct translation with minimal adjustment effort. 

\subsection{Threats to Validity}
One threat lies in the construction quality of \kg{}. We sample part of the code-dependency graph, pointer-usage information, and Rust-oriented annotations, and manually verify their correctness and accuracy to eliminate implementation error of KG construction.  
Another threat is potential performance degradation of the translated Rust projects compared to original C projects. We assess this by measuring the runtime performance of four projects that are successfully and correctly translated by \ourtool{} (\ie{} \textit{avl}, \textit{rgba}, \textit{ht}, \textit{bst}) and comparing them with their C versions. We do not observe any  performance degradation, \eg{} the translated Rust project runs 54\% faster than the C version for the project \textit{bst}.


\section{Conclusion}
In this paper, we present \ourtool{}, a project-level C-to-Rust translation framework that leverages a novel \kg{} and its synergy with LLMs. Our KG captures code-unit dependencies, global pointer-usage information, and Rust-oriented annotations, providing rich context for LLM-guided translation. Experiments on real-world C projects show that \ourtool{} substantially improves code safety and correctness, highlighting a promising direction for automated, safe, and idiomatic C-to-Rust translation.

\section{Data Availability}
Our data and code are included in our replication package~\cite{KGC2Rust}.

\begin{acks}
This work is supported by the National Natural Science Foundation of China (Grant No.~62332005). 
\end{acks}

\balance
\bibliographystyle{ACM-Reference-Format} 

\bibliography{ref}

@inproceedings{fan2020ac,
  title={AC/C++ code vulnerability dataset with code changes and CVE summaries},
  author={Fan, Jiahao and Li, Yi and Wang, Shaohua and Nguyen, Tien N},
  booktitle={Proceedings of the 17th international conference on mining software repositories},
  pages={508--512},
  year={2020}
}

@inproceedings{matsakis2014rust,
  title={The rust language},
  author={Matsakis, Nicholas D and Klock, Felix S},
  booktitle={Proceedings of the 2014 ACM SIGAda annual conference on High integrity language technology},
  pages={103--104},
  year={2014}
}

@article{DBLP:journals/pieee/SaltzerS75,
  author       = {Jerome H. Saltzer and
                  Michael D. Schroeder},
  title        = {The protection of information in computer systems},
  journal      = {Proc. {IEEE}},
  volume       = {63},
  number       = {9},
  pages        = {1278--1308},
  year         = {1975},
  url          = {https://doi.org/10.1109/PROC.1975.9939},
  doi          = {10.1109/PROC.1975.9939},
  bibsource    = {dblp computer science bibliography, https://dblp.org}
}

@inproceedings{crown,
  author       = {Hanliang Zhang and
                  Cristina David and
                  Yijun Yu and
                  Meng Wang},
  title        = {Ownership Guided {C} to Rust Translation},
  booktitle    = {Computer Aided Verification - 35th International Conference, {CAV}
                  2023, Paris, France, July 17-22, 2023, Proceedings, Part {III}},
  series       = {Lecture Notes in Computer Science},
  volume       = {13966},
  pages        = {459--482},
  publisher    = {Springer},
  year         = {2023},
  url          = {https://doi.org/10.1007/978-3-031-37709-9\_22},
  doi          = {10.1007/978-3-031-37709-9\_22},
  bibsource    = {dblp computer science bibliography, https://dblp.org}
}

@article{DBLP:journals/siamcomp/Tarjan72,
  author       = {Robert Endre Tarjan},
  title        = {Depth-First Search and Linear Graph Algorithms},
  journal      = {{SIAM} J. Comput.},
  volume       = {1},
  number       = {2},
  pages        = {146--160},
  year         = {1972},
  url          = {https://doi.org/10.1137/0201010},
  doi          = {10.1137/0201010},
  bibsource    = {dblp computer science bibliography, https://dblp.org}
}

@article{PR2,
  author       = {Yifei Gao and
                  Chengpeng Wang and
                  Pengxiang Huang and
                  Xuwei Liu and
                  Mingwei Zheng and
                  Xiangyu Zhang},
  title        = {{PR2:} Peephole Raw Pointer Rewriting with LLMs for Translating {C}
                  to Safer Rust},
  journal      = {CoRR},
  volume       = {abs/2505.04852},
  year         = {2025},
  url          = {https://doi.org/10.48550/arXiv.2505.04852},
  doi          = {10.48550/ARXIV.2505.04852},
  eprinttype    = {arXiv},
  eprint       = {2505.04852},
  bibsource    = {dblp computer science bibliography, https://dblp.org}
}

@article{DBLP:journals/corr/abs-2501-14257,
  author       = {Vikram Nitin and
                  Rahul Krishna and
                  Luiz Lemos do Valle and
                  Baishakhi Ray},
  title        = {C2SaferRust: Transforming {C} Projects into Safer Rust with NeuroSymbolic
                  Techniques},
  journal      = {CoRR},
  volume       = {abs/2501.14257},
  year         = {2025},
  url          = {https://doi.org/10.48550/arXiv.2501.14257},
  doi          = {10.48550/ARXIV.2501.14257},
  eprinttype    = {arXiv},
  eprint       = {2501.14257},
  bibsource    = {dblp computer science bibliography, https://dblp.org}
}

@article{DBLP:journals/corr/abs-2405-11514,
  author       = {Hasan Ferit Eniser and
                  Hanliang Zhang and
                  Cristina David and
                  Meng Wang and
                  Maria Christakis and
                  Brandon Paulsen and
                  Joey Dodds and
                  Daniel Kroening},
  title        = {Towards Translating Real-World Code with LLMs: {A} Study of Translating
                  to Rust},
  journal      = {CoRR},
  volume       = {abs/2405.11514},
  year         = {2024},
  url          = {https://doi.org/10.48550/arXiv.2405.11514},
  doi          = {10.48550/ARXIV.2405.11514},
  eprinttype    = {arXiv},
  eprint       = {2405.11514},
  bibsource    = {dblp computer science bibliography, https://dblp.org}
}

@article{DBLP:journals/corr/abs-2503-17741,
  author       = {Xuemeng Cai and
                  Jiakun Liu and
                  Xiping Huang and
                  Yijun Yu and
                  Haitao Wu and
                  Chunmiao Li and
                  Bo Wang and
                  Imam Nur Bani Yusuf and
                  Lingxiao Jiang},
  title        = {RustMap: Towards Project-Scale C-to-Rust Migration via Program Analysis
                  and {LLM}},
  journal      = {CoRR},
  volume       = {abs/2503.17741},
  year         = {2025},
  url          = {https://doi.org/10.48550/arXiv.2503.17741},
  doi          = {10.48550/ARXIV.2503.17741},
  eprinttype    = {arXiv},
  eprint       = {2503.17741},
  bibsource    = {dblp computer science bibliography, https://dblp.org}
}

@article{DBLP:journals/corr/abs-2404-18852,
  author       = {Aidan Z. H. Yang and
                  Yoshiki Takashima and
                  Brandon Paulsen and
                  Josiah Dodds and
                  Daniel Kroening},
  title        = {{VERT:} Verified Equivalent Rust Transpilation with Few-Shot Learning},
  journal      = {CoRR},
  volume       = {abs/2404.18852},
  year         = {2024},
  url          = {https://doi.org/10.48550/arXiv.2404.18852},
  doi          = {10.48550/ARXIV.2404.18852},
  eprinttype    = {arXiv},
  eprint       = {2404.18852},
  bibsource    = {dblp computer science bibliography, https://dblp.org}
}

@inproceedings{DBLP:conf/icse/PanIKSWMSPSJ24,
  author       = {Rangeet Pan and
                  Ali Reza Ibrahimzada and
                  Rahul Krishna and
                  Divya Sankar and
                  Lambert Pouguem Wassi and
                  Michele Merler and
                  Boris Sobolev and
                  Raju Pavuluri and
                  Saurabh Sinha and
                  Reyhaneh Jabbarvand},
  title        = {Lost in Translation: {A} Study of Bugs Introduced by Large Language
                  Models while Translating Code},
  booktitle    = {Proceedings of the 46th {IEEE/ACM} International Conference on Software
                  Engineering, {ICSE} 2024, Lisbon, Portugal, April 14-20, 2024},
  pages        = {82:1--82:13},
  publisher    = {{ACM}},
  year         = {2024},
  url          = {https://doi.org/10.1145/3597503.3639226},
  doi          = {10.1145/3597503.3639226},
  bibsource    = {dblp computer science bibliography, https://dblp.org}
}

@article{DBLP:journals/corr/abs-2412-14234,
  author       = {Manish Shetty and
                  Naman Jain and
                  Adwait Godbole and
                  Sanjit A. Seshia and
                  Koushik Sen},
  title        = {Syzygy: Dual Code-Test {C} to (safe) Rust Translation using LLMs and
                  Dynamic Analysis},
  journal      = {CoRR},
  volume       = {abs/2412.14234},
  year         = {2024},
  url          = {https://doi.org/10.48550/arXiv.2412.14234},
  doi          = {10.48550/ARXIV.2412.14234},
  eprinttype    = {arXiv},
  eprint       = {2412.14234},
  bibsource    = {dblp computer science bibliography, https://dblp.org}
}

@article{DBLP:journals/corr/abs-2503-12511,
  author       = {Tianyang Zhou and
                  Haowen Lin and
                  Somesh Jha and
                  Mihai Christodorescu and
                  Kirill Levchenko and
                  Varun Chandrasekaran},
  title        = {LLM-Driven Multi-step Translation from {C} to Rust using Static Analysis},
  journal      = {CoRR},
  volume       = {abs/2503.12511},
  year         = {2025},
  url          = {https://doi.org/10.48550/arXiv.2503.12511},
  doi          = {10.48550/ARXIV.2503.12511},
  eprinttype    = {arXiv},
  eprint       = {2503.12511},
  bibsource    = {dblp computer science bibliography, https://dblp.org}
}

@article{DBLP:journals/corr/abs-2503-03698,
  author       = {Shahram Esmaeilsabzali and
                  Arayi Khalatyan and
                  Zhijun Mo and
                  Sruthi Venkatanarayanan and
                  Shengjie Xu},
  title        = {{AEGIS:} Towards Formalized and Practical Memory-Safe Execution of
                  {C} programs via {MSWASM}},
  journal      = {CoRR},
  volume       = {abs/2503.03698},
  year         = {2025},
  url          = {https://doi.org/10.48550/arXiv.2503.03698},
  doi          = {10.48550/ARXIV.2503.03698},
  eprinttype    = {arXiv},
  eprint       = {2503.03698},
  bibsource    = {dblp computer science bibliography, https://dblp.org}
}

@article{yin2024safemd,
  title={SafeMD: Ownership-Based Safe Memory Deallocation for C Programs},
  author={Yin, Xiaohua and Huang, Zhiqiu and Kan, Shuanglong and Shen, Guohua},
  year={2024},
  publisher={Preprints}
}

@article{DBLP:journals/ieeesp/Larsen24,
  author       = {Per Larsen},
  title        = {Migrating {C} to Rust for Memory Safety},
  journal      = {{IEEE} Secur. Priv.},
  volume       = {22},
  number       = {4},
  pages        = {22--29},
  year         = {2024},
  url          = {https://doi.org/10.1109/MSEC.2024.3385357},
  doi          = {10.1109/MSEC.2024.3385357},
  bibsource    = {dblp computer science bibliography, https://dblp.org}
}

@article{DBLP:journals/corr/abs-2409-10506,
  author       = {Momoko Shiraishi and
                  Takahiro Shinagawa},
  title        = {Context-aware Code Segmentation for C-to-Rust Translation using Large
                  Language Models},
  journal      = {CoRR},
  volume       = {abs/2409.10506},
  year         = {2024},
  url          = {https://doi.org/10.48550/arXiv.2409.10506},
  doi          = {10.48550/ARXIV.2409.10506},
  eprinttype    = {arXiv},
  eprint       = {2409.10506},
  bibsource    = {dblp computer science bibliography, https://dblp.org}
}

@article{DBLP:journals/ese/HongR25,
  author       = {Jaemin Hong and
                  Sukyoung Ryu},
  title        = {Type-migrating C-to-Rust translation using a large language model},
  journal      = {Empir. Softw. Eng.},
  volume       = {30},
  number       = {1},
  pages        = {3},
  year         = {2025},
  url          = {https://doi.org/10.1007/s10664-024-10573-2},
  doi          = {10.1007/S10664-024-10573-2},
  bibsource    = {dblp computer science bibliography, https://dblp.org}
}

@article{DBLP:journals/corr/abs-2505-10708,
  author       = {Muhammad Farrukh and
                  Smeet Shah and
                  Baris Coskun and
                  Michalis Polychronakis},
  title        = {SafeTrans: LLM-assisted Tran2spilation from {C} to Rust},
  journal      = {CoRR},
  volume       = {abs/2505.10708},
  year         = {2025},
  url          = {https://doi.org/10.48550/arXiv.2505.10708},
  doi          = {10.48550/ARXIV.2505.10708},
  eprinttype    = {arXiv},
  eprint       = {2505.10708},
  bibsource    = {dblp computer science bibliography, https://dblp.org}
}

@inproceedings{DBLP:conf/icse/LingYWWCH22,
  author       = {Michael Ling and
                  Yijun Yu and
                  Haitao Wu and
                  Yuan Wang and
                  James R. Cordy and
                  Ahmed E. Hassan},
  title        = {In Rust We Trust - {A} Transpiler from Unsafe {C} to Safer Rust},
  booktitle    = {44th {IEEE/ACM} International Conference on Software Engineering:
                  Companion Proceedings, {ICSE} Companion 2022, Pittsburgh, PA, USA,
                  May 22-24, 2022},
  pages        = {354--355},
  publisher    = {{ACM/IEEE}},
  year         = {2022},
  url          = {https://doi.org/10.1145/3510454.3528640},
  doi          = {10.1145/3510454.3528640},
  bibsource    = {dblp computer science bibliography, https://dblp.org}
}

@article{DBLP:journals/pacmpl/EmreBPSDH23,
  author       = {Mehmet Emre and
                  Peter Boyland and
                  Aesha Parekh and
                  Ryan Schroeder and
                  Kyle Dewey and
                  Ben Hardekopf},
  title        = {Aliasing Limits on Translating {C} to Safe Rust},
  journal      = {Proc. {ACM} Program. Lang.},
  volume       = {7},
  number       = {{OOPSLA1}},
  pages        = {551--579},
  year         = {2023},
  url          = {https://doi.org/10.1145/3586046},
  doi          = {10.1145/3586046},
  bibsource    = {dblp computer science bibliography, https://dblp.org}
}

@article{DBLP:journals/pacmpl/EmreSDH21,
  author       = {Mehmet Emre and
                  Ryan Schroeder and
                  Kyle Dewey and
                  Ben Hardekopf},
  title        = {Translating {C} to safer Rust},
  journal      = {Proc. {ACM} Program. Lang.},
  volume       = {5},
  number       = {{OOPSLA}},
  pages        = {1--29},
  year         = {2021},
  url          = {https://doi.org/10.1145/3485498},
  doi          = {10.1145/3485498},
  bibsource    = {dblp computer science bibliography, https://dblp.org}
}

@article{DBLP:journals/pacmpl/HongR24,
  author       = {Jaemin Hong and
                  Sukyoung Ryu},
  title        = {Don't Write, but Return: Replacing Output Parameters with Algebraic
                  Data Types in C-to-Rust Translation},
  journal      = {Proc. {ACM} Program. Lang.},
  volume       = {8},
  number       = {{PLDI}},
  pages        = {716--740},
  year         = {2024},
  url          = {https://doi.org/10.1145/3656406},
  doi          = {10.1145/3656406},
  bibsource    = {dblp computer science bibliography, https://dblp.org}
}

@inproceedings{DBLP:conf/kbse/HongR24,
  author       = {Jaemin Hong and
                  Sukyoung Ryu},
  editor       = {Vladimir Filkov and
                  Baishakhi Ray and
                  Minghui Zhou},
  title        = {To Tag, or Not to Tag: Translating C's Unions to Rust's Tagged Unions},
  booktitle    = {Proceedings of the 39th {IEEE/ACM} International Conference on Automated
                  Software Engineering, {ASE} 2024, Sacramento, CA, USA, October 27
                  - November 1, 2024},
  pages        = {40--52},
  publisher    = {{ACM}},
  year         = {2024},
  url          = {https://doi.org/10.1145/3691620.3694985},
  doi          = {10.1145/3691620.3694985},
  bibsource    = {dblp computer science bibliography, https://dblp.org}
}

@article{DBLP:journals/corr/abs-2412-15042,
  author       = {Aymeric Fromherz and
                  Jonathan Protzenko},
  title        = {Compiling {C} to Safe Rust, Formalized},
  journal      = {CoRR},
  volume       = {abs/2412.15042},
  year         = {2024},
  url          = {https://doi.org/10.48550/arXiv.2412.15042},
  doi          = {10.48550/ARXIV.2412.15042},
  eprinttype    = {arXiv},
  eprint       = {2412.15042},
  bibsource    = {dblp computer science bibliography, https://dblp.org}
}

@article{DBLP:journals/corr/abs-2506-01427,
  author       = {Jaemin Hong and
                  Sukyoung Ryu},
  title        = {Forcrat: Automatic {I/O} {API} Translation from {C} to Rust via Origin
                  and Capability Analysis},
  journal      = {CoRR},
  volume       = {abs/2506.01427},
  year         = {2025},
  url          = {https://doi.org/10.48550/arXiv.2506.01427},
  doi          = {10.48550/ARXIV.2506.01427},
  eprinttype    = {arXiv},
  eprint       = {2506.01427},
  bibsource    = {dblp computer science bibliography, https://dblp.org}
}

@inproceedings{DBLP:conf/ndss/LiWLSK25,
  author       = {Ruishi Li and
                  Bo Wang and
                  Tianyu Li and
                  Prateek Saxena and
                  Ashish Kundu},
  title        = {Translating {C} To Rust: Lessons from a User Study},
  booktitle    = {32nd Annual Network and Distributed System Security Symposium, {NDSS}
                  2025, San Diego, California, USA, February 24-28, 2025},
  publisher    = {The Internet Society},
  year         = {2025},
  url          = {https://www.ndss-symposium.org/ndss-paper/translating-c-to-rust-lessons-from-a-user-study/},
  bibsource    = {dblp computer science bibliography, https://dblp.org}
}

@article{DBLP:journals/corr/abs-2405-18574,
  author       = {Vikram Nitin and
                  Baishakhi Ray},
  title        = {SpecTra: Enhancing the Code Translation Ability of Language Models
                  by Generating Multi-Modal Specifications},
  journal      = {CoRR},
  volume       = {abs/2405.18574},
  year         = {2024},
  url          = {https://doi.org/10.48550/arXiv.2405.18574},
  doi          = {10.48550/ARXIV.2405.18574},
  eprinttype    = {arXiv},
  eprint       = {2405.18574},
  bibsource    = {dblp computer science bibliography, https://dblp.org}
}

@article{DBLP:journals/corr/abs-2505-15858,
  author       = {HoHyun Sim and
                  Hyeonjoong Cho and
                  Yeonghyeon Go and
                  Zhoulai Fu and
                  Ali Shokri and
                  Binoy Ravindran},
  title        = {Large Language Model-Powered Agent for {C} to Rust Code Translation},
  journal      = {CoRR},
  volume       = {abs/2505.15858},
  year         = {2025},
  url          = {https://doi.org/10.48550/arXiv.2505.15858},
  doi          = {10.48550/ARXIV.2505.15858},
  eprinttype    = {arXiv},
  eprint       = {2505.15858},
  bibsource    = {dblp computer science bibliography, https://dblp.org}
}

@article{DBLP:journals/corr/abs-2504-15254,
  author       = {Anirudh Khatry and
                  Robert Zhang and
                  Jia Pan and
                  Ziteng Wang and
                  Qiaochu Chen and
                  Greg Durrett and
                  Isil Dillig},
  title        = {CRUST-Bench: {A} Comprehensive Benchmark for C-to-safe-Rust Transpilation},
  journal      = {CoRR},
  volume       = {abs/2504.15254},
  year         = {2025},
  url          = {https://doi.org/10.48550/arXiv.2504.15254},
  doi          = {10.48550/ARXIV.2504.15254},
  eprinttype    = {arXiv},
  eprint       = {2504.15254},
  bibsource    = {dblp computer science bibliography, https://dblp.org}
}

@article{DBLP:journals/corr/abs-2410-24117,
  author       = {Ali Reza Ibrahimzada and
                  Kaiyao Ke and
                  Mrigank Pawagi and
                  Muhammad Salman Abid and
                  Rangeet Pan and
                  Saurabh Sinha and
                  Reyhaneh Jabbarvand},
  title        = {AlphaTrans: {A} Neuro-Symbolic Compositional Approach for Repository-Level
                  Code Translation and Validation},
  journal      = {Proc. {ACM} Softw. Eng.},
  volume       = {2},
  number       = {{FSE}},
  pages        = {2454--2476},
  year         = {2025},
  url          = {https://doi.org/10.1145/3729379},
  doi          = {10.1145/3729379},
  bibsource    = {dblp computer science bibliography, https://dblp.org}
}

@article{DBLP:journals/corr/abs-2503-18305,
  author       = {Guangsheng Ou and
                  Mingwei Liu and
                  Yuxuan Chen and
                  Xueying Du and
                  Shengbo Wang and
                  Zekai Zhang and
                  Xin Peng and
                  Zibin Zheng},
  title        = {Enhancing LLM-based Code Translation in Repository Context via Triple
                  Knowledge-Augmented},
  journal      = {CoRR},
  volume       = {abs/2503.18305},
  year         = {2025},
  url          = {https://doi.org/10.48550/arXiv.2503.18305},
  doi          = {10.48550/ARXIV.2503.18305},
  eprinttype    = {arXiv},
  eprint       = {2503.18305},
  bibsource    = {dblp computer science bibliography, https://dblp.org}
}

@article{DBLP:journals/corr/abs-2411-13990,
  author       = {Guangsheng Ou and
                  Mingwei Liu and
                  Yuxuan Chen and
                  Xin Peng and
                  Zibin Zheng},
  title        = {Repository-level Code Translation Benchmark Targeting Rust},
  journal      = {CoRR},
  volume       = {abs/2411.13990},
  year         = {2024},
  url          = {https://doi.org/10.48550/arXiv.2411.13990},
  doi          = {10.48550/ARXIV.2411.13990},
  eprinttype    = {arXiv},
  eprint       = {2411.13990},
  bibsource    = {dblp computer science bibliography, https://dblp.org}
}

@article{DBLP:journals/corr/abs-2409-19894,
  author       = {Zhiqiang Yuan and
                  Weitong Chen and
                  Hanlin Wang and
                  Kai Yu and
                  Xin Peng and
                  Yiling Lou},
  title        = {{TRANSAGENT:} An LLM-Based Multi-Agent System for Code Translation},
  journal      = {CoRR},
  volume       = {abs/2409.19894},
  year         = {2024},
  url          = {https://doi.org/10.48550/arXiv.2409.19894},
  doi          = {10.48550/ARXIV.2409.19894},
  eprinttype    = {arXiv},
  eprint       = {2409.19894},
  bibsource    = {dblp computer science bibliography, https://dblp.org}
}

@article{DBLP:journals/corr/abs-2308-01240,
  author       = {Zhiqiang Yuan and
                  Junwei Liu and
                  Qiancheng Zi and
                  Mingwei Liu and
                  Xin Peng and
                  Yiling Lou},
  title        = {Evaluating Instruction-Tuned Large Language Models on Code Comprehension
                  and Generation},
  journal      = {CoRR},
  volume       = {abs/2308.01240},
  year         = {2023},
  url          = {https://doi.org/10.48550/arXiv.2308.01240},
  doi          = {10.48550/ARXIV.2308.01240},
  eprinttype    = {arXiv},
  eprint       = {2308.01240},
  bibsource    = {dblp computer science bibliography, https://dblp.org}
}

@inproceedings{DBLP:conf/icse/XiaWZ23,
  author       = {Chunqiu Steven Xia and
                  Yuxiang Wei and
                  Lingming Zhang},
  title        = {Automated Program Repair in the Era of Large Pre-trained Language
                  Models},
  booktitle    = {45th {IEEE/ACM} International Conference on Software Engineering,
                  {ICSE} 2023, Melbourne, Australia, May 14-20, 2023},
  pages        = {1482--1494},
  publisher    = {{IEEE}},
  year         = {2023},
  url          = {https://doi.org/10.1109/ICSE48619.2023.00129},
  doi          = {10.1109/ICSE48619.2023.00129},
  bibsource    = {dblp computer science bibliography, https://dblp.org}
}

@article{DBLP:journals/pacmse/Yuan0DW00L24,
  author       = {Zhiqiang Yuan and
                  Mingwei Liu and
                  Shiji Ding and
                  Kaixin Wang and
                  Yixuan Chen and
                  Xin Peng and
                  Yiling Lou},
  title        = {Evaluating and Improving ChatGPT for Unit Test Generation},
  journal      = {Proc. {ACM} Softw. Eng.},
  volume       = {1},
  number       = {{FSE}},
  pages        = {1703--1726},
  year         = {2024},
  url          = {https://doi.org/10.1145/3660783},
  doi          = {10.1145/3660783},
  bibsource    = {dblp computer science bibliography, https://dblp.org}
}

@inproceedings{DBLP:conf/icse/BouzeniaDP25,
  author       = {Islem Bouzenia and
                  Premkumar T. Devanbu and
                  Michael Pradel},
  title        = {RepairAgent: An Autonomous, LLM-Based Agent for Program Repair},
  booktitle    = {47th {IEEE/ACM} International Conference on Software Engineering,
                  {ICSE} 2025, Ottawa, ON, Canada, April 26 - May 6, 2025},
  pages        = {2188--2200},
  publisher    = {{IEEE}},
  year         = {2025},
  url          = {https://doi.org/10.1109/ICSE55347.2025.00157},
  doi          = {10.1109/ICSE55347.2025.00157},
  bibsource    = {dblp computer science bibliography, https://dblp.org}
}

@article{DBLP:journals/pacmse/Yang0YK0LHMJ024,
  author       = {Zhen Yang and
                  Fang Liu and
                  Zhongxing Yu and
                  Jacky Wai Keung and
                  Jia Li and
                  Shuo Liu and
                  Yifan Hong and
                  Xiaoxue Ma and
                  Zhi Jin andns
                  Ge Li},
  title        = {Exploring and Unleashing the Power of Large Language Models in Automated
                  Code Translation},
  journal      = {Proc. {ACM} Softw. Eng.},
  volume       = {1},
  number       = {{FSE}},
  pages        = {1585--1608},
  year         = {2024},
  url          = {https://doi.org/10.1145/3660778},
  doi          = {10.1145/3660778},
  bibsource    = {dblp computer science bibliography, https://dblp.org}
}

@article{DBLP:journals/tosem/DongJJL24,
  author       = {Yihong Dong and
                  Xue Jiang and
                  Zhi Jin and
                  Ge Li},
  title        = {Self-Collaboration Code Generation via ChatGPT},
  journal      = {{ACM} Trans. Softw. Eng. Methodol.},
  volume       = {33},
  number       = {7},
  pages        = {189:1--189:38},
  year         = {2024},
  url          = {https://doi.org/10.1145/3672459},
  doi          = {10.1145/3672459},
  bibsource    = {dblp computer science bibliography, https://dblp.org}
}

@article{DBLP:journals/tosem/XuCSZL22,
  author       = {Hui Xu and
                  Zhuangbin Chen and
                  Mingshen Sun and
                  Yangfan Zhou and
                  Michael R. Lyu},
  title        = {Memory-Safety Challenge Considered Solved? An In-Depth Study with
                  All Rust CVEs},
  journal      = {{ACM} Trans. Softw. Eng. Methodol.},
  volume       = {31},
  number       = {1},
  pages        = {3:1--3:25},
  year         = {2022},
  url          = {https://doi.org/10.1145/3466642},
  doi          = {10.1145/3466642},
  bibsource    = {dblp computer science bibliography, https://dblp.org}
}

@inproceedings{DBLP:conf/sigsoft/ZhangKPX23,
  author       = {Yuchen Zhang and
                  Ashish Kundu and
                  Georgios Portokalidis and
                  Jun Xu},
  editor       = {Satish Chandra and
                  Kelly Blincoe and
                  Paolo Tonella},
  title        = {On the Dual Nature of Necessity in Use of Rust Unsafe Code},
  booktitle    = {Proceedings of the 31st {ACM} Joint European Software Engineering
                  Conference and Symposium on the Foundations of Software Engineering,
                  {ESEC/FSE} 2023, San Francisco, CA, USA, December 3-9, 2023},
  pages        = {2032--2037},
  publisher    = {{ACM}},
  year         = {2023},
  url          = {https://doi.org/10.1145/3611643.3613878},
  doi          = {10.1145/3611643.3613878},
  bibsource    = {dblp computer science bibliography, https://dblp.org}
}

@article{DBLP:journals/tosem/ZhengWZCL24,
  author       = {Xiaoye Zheng and
                  Zhiyuan Wan and
                  Yun Zhang and
                  Rui Chang and
                  David Lo},
  title        = {A Closer Look at the Security Risks in the Rust Ecosystem},
  journal      = {{ACM} Trans. Softw. Eng. Methodol.},
  volume       = {33},
  number       = {2},
  pages        = {34:1--34:30},
  year         = {2024},
  url          = {https://doi.org/10.1145/3624738},
  doi          = {10.1145/3624738},
  bibsource    = {dblp computer science bibliography, https://dblp.org}
}

@article{DBLP:journals/corr/abs-2201-07845,
  author       = {Victor Yodaiken},
  title        = {How {ISO} {C} became unusable for operating systems development},
  journal      = {CoRR},
  volume       = {abs/2201.07845},
  year         = {2022},
  url          = {https://arxiv.org/abs/2201.07845},
  eprinttype    = {arXiv},
  eprint       = {2201.07845},
  bibsource    = {dblp computer science bibliography, https://dblp.org}
}

@article{DBLP:journals/corr/PatelR13a,
  author       = {Rajendra Patel and
                  Arvind Rajawat},
  title        = {A Survey of Embedded Software Profiling Methodologies},
  journal      = {CoRR},
  volume       = {abs/1312.2949},
  year         = {2013},
  url          = {http://arxiv.org/abs/1312.2949},
  eprinttype    = {arXiv},
  eprint       = {1312.2949},
  bibsource    = {dblp computer science bibliography, https://dblp.org}
}

@inproceedings{DBLP:conf/icse/FanGMRT23,
  author       = {Zhiyu Fan and
                  Xiang Gao and
                  Martin Mirchev and
                  Abhik Roychoudhury and
                  Shin Hwei Tan},
  title        = {Automated Repair of Programs from Large Language Models},
  booktitle    = {45th {IEEE/ACM} International Conference on Software Engineering,
                  {ICSE} 2023, Melbourne, Australia, May 14-20, 2023},
  pages        = {1469--1481},
  publisher    = {{IEEE}},
  year         = {2023},
  url          = {https://doi.org/10.1109/ICSE48619.2023.00128},
  doi          = {10.1109/ICSE48619.2023.00128},
  bibsource    = {dblp computer science bibliography, https://dblp.org}
}

@article{DBLP:journals/corr/abs-2412-08035,
  author       = {Hanliang Zhang and
                  Cristina David and
                  Meng Wang and
                  Brandon Paulsen and
                  Daniel Kroening},
  title        = {Scalable, Validated Code Translation of Entire Projects using Large
                  Language Models},
  journal      = {CoRR},
  volume       = {abs/2412.08035},
  year         = {2024},
  url          = {https://doi.org/10.48550/arXiv.2412.08035},
  doi          = {10.48550/ARXIV.2412.08035},
  eprinttype    = {arXiv},
  eprint       = {2412.08035},
  bibsource    = {dblp computer science bibliography, https://dblp.org}
}

@article{DBLP:journals/corr/abs-2009-10297,
  author       = {Shuo Ren and
                  Daya Guo and
                  Shuai Lu and
                  Long Zhou and
                  Shujie Liu and
                  Duyu Tang and
                  Neel Sundaresan and
                  Ming Zhou and
                  Ambrosio Blanco and
                  Shuai Ma},
  title        = {CodeBLEU: a Method for Automatic Evaluation of Code Synthesis},
  journal      = {CoRR},
  volume       = {abs/2009.10297},
  year         = {2020},
  url          = {https://arxiv.org/abs/2009.10297},
  eprinttype    = {arXiv},
  eprint       = {2009.10297},
  bibsource    = {dblp computer science bibliography, https://dblp.org}
}

@inproceedings{DBLP:conf/emnlp/YanTLCW23,
  author       = {Weixiang Yan and
                  Yuchen Tian and
                  Yunzhe Li and
                  Qian Chen and
                  Wen Wang},
  editor       = {Houda Bouamor and
                  Juan Pino and
                  Kalika Bali},
  title        = {CodeTransOcean: {A} Comprehensive Multilingual Benchmark for Code
                  Translation},
  booktitle    = {Findings of the Association for Computational Linguistics: {EMNLP}
                  2023, Singapore, December 6-10, 2023},
  pages        = {5067--5089},
  publisher    = {Association for Computational Linguistics},
  year         = {2023},
  url          = {https://doi.org/10.18653/v1/2023.findings-emnlp.337},
  doi          = {10.18653/V1/2023.FINDINGS-EMNLP.337},
  bibsource    = {dblp computer science bibliography, https://dblp.org}
}

@article{DBLP:journals/access/AlmanasraS25,
  author       = {Sally Almanasra and
                  Khaled Suwais},
  title        = {Analysis of ChatGPT-Generated Codes Across Multiple Programming Languages},
  journal      = {{IEEE} Access},
  volume       = {13},
  pages        = {23580--23596},
  year         = {2025},
  url          = {https://doi.org/10.1109/ACCESS.2025.3538050},
  doi          = {10.1109/ACCESS.2025.3538050},
  bibsource    = {dblp computer science bibliography, https://dblp.org}
}

@article{DBLP:journals/corr/abs-2502-07399,
  author       = {Rundong Liu and
                  Andre Frade and
                  Amal Vaidya and
                  Maxime Labonne and
                  Marcus Kaiser and
                  Bismayan Chakrabarti and
                  Jonathan Budd and
                  Sean J. Moran},
  title        = {On Iterative Evaluation and Enhancement of Code Quality Using GPT-4o},
  journal      = {CoRR},
  volume       = {abs/2502.07399},
  year         = {2025},
  url          = {https://doi.org/10.48550/arXiv.2502.07399},
  doi          = {10.48550/ARXIV.2502.07399},
  eprinttype    = {arXiv},
  eprint       = {2502.07399},
  bibsource    = {dblp computer science bibliography, https://dblp.org}
}

@INPROCEEDINGS{11061239,
  author={Mashaal, Abdelrahman and Helmy, Omar and Ashor, Omar and Mahmoud, Amira T. and WalaaMedhat},
  booktitle={2024 12th International Japan-Africa Conference on Electronics, Communications, and Computations (JAC-ECC)}, 
  title={Evaluation of GPT 4o for Mobile Applications Code Conversion}, 
  year={2024},
  volume={},
  number={},
  pages={219-224},
  doi={10.1109/JAC-ECC64419.2024.11061239}}

@article{DBLP:journals/corr/abs-2412-17744,
  author       = {Yanli Wang and
                  Yanlin Wang and
                  Suiquan Wang and
                  Daya Guo and
                  Jiachi Chen and
                  John C. Grundy and
                  Xilin Liu and
                  Yuchi Ma and
                  Mingzhi Mao and
                  Hongyu Zhang and
                  Zibin Zheng},
  title        = {RepoTransBench: {A} Real-World Benchmark for Repository-Level Code
                  Translation},
  journal      = {CoRR},
  volume       = {abs/2412.17744},
  year         = {2024},
  url          = {https://doi.org/10.48550/arXiv.2412.17744},
  doi          = {10.48550/ARXIV.2412.17744},
  eprinttype    = {arXiv},
  eprint       = {2412.17744},
  bibsource    = {dblp computer science bibliography, https://dblp.org}
}

@misc{guan2025repotransagentmultiagentllmframework,
      title={RepoTransAgent: Multi-Agent LLM Framework for Repository-Aware Code Translation}, 
      author={Ziqi Guan and Xin Yin and Zhiyuan Peng and Chao Ni},
      year={2025},
      eprint={2508.17720},
      archivePrefix={arXiv},
      primaryClass={cs.SE},
      url={https://arxiv.org/abs/2508.17720}, 
}

@INPROCEEDINGS{10938485,
  author={Bhattarai, Manish and Santos, Javier E. and Jones, Shawn and Biswas, Ayan and Alexandrov, Boian and O'Malley, Daniel},
  booktitle={2024 IEEE High Performance Extreme Computing Conference (HPEC)}, 
  title={Enhancing Code Translation in Language Models with Few-Shot Learning via Retrieval-Augmented Generation}, 
  year={2024},
  volume={},
  number={},
  pages={1-8},
  doi={10.1109/HPEC62836.2024.10938485}}

@inproceedings{DBLP:conf/nips/RoziereLCL20,
  author       = {Baptiste Rozi{\`{e}}re and
                  Marie{-}Anne Lachaux and
                  Lowik Chanussot and
                  Guillaume Lample},
  title        = {Unsupervised Translation of Programming Languages},
  booktitle    = {Advances in Neural Information Processing Systems 33: Annual Conference
                  on Neural Information Processing Systems 2020, NeurIPS 2020, December
                  6-12, 2020, virtual},
  year         = {2020},
}

@INPROCEEDINGS{11126570,
  author={Zhou, Han and Luo, Yu and Zhang, Mengtao and Xu, Dianxiang},
  booktitle={2025 IEEE 49th Annual Computers, Software, and Applications Conference (COMPSAC)}, 
  title={C2RustTV: An LLM-based Framework for C to Rust Translation and Validation}, 
  year={2025},
  volume={},
  number={},
  pages={1254-1259},
  doi={10.1109/COMPSAC65507.2025.00158}}

@article{DBLP:journals/cacm/HongR25,
  author       = {Jaemin Hong and
                  Sukyoung Ryu},
  title        = {Automatically Translating {C} to Rust},
  journal      = {Commun. {ACM}},
  volume       = {68},
  number       = {11},
  pages        = {58--65},
  year         = {2025},
  url          = {https://doi.org/10.1145/3737696},
  doi          = {10.1145/3737696},
  bibsource    = {dblp computer science bibliography, https://dblp.org}
}

@article{shiraishitoward,
  title={Toward LLM-based Large-scale C-to-Rust Code Translation},
  author={Shiraishi, Momoko and Shinagawa, Takahiro}
}

@article{c2rust,
  url={https://github.com/immunant/c2rust},
  author={c2rust},
  year={2025}
}

@article{citrus,
  url={https://gitlab.com/citrus-rs/citrus#citrus-convert-c-to-rust},
  author={citrus},
  year={2025}
}

@article{Corrode,
  url={https://github.com/jameysharp/corrode},
  author={Corrode},
  year={2017}
}

@article{Rust4Linux,
  url={https://github.com/Rust-for-Linux/linux},
  author={Rust-for-Linux},
}

@article{quadtree,
  url={https://github.com/thejefflarson/quadtree},
  author={Quadtree},
}

@article{freeFunc,
  url={https://github.com/thejefflarson/quadtree/blob/master/src/point.c#L14},
  author={quadtree\_point\_free},
}

@misc{hanley2023rust,
  title={Rust Won’t Save Us: An Analysis of 2023’s Known Exploited Vulnerabilities},
  author={Hanley, Zach},
  year={2023}
}

@article{libtree,
  url={https://github.com/haampie/libtree/blob/master/libtree.c#L191},
  author={libtree},
}

@article{chatgpt,
  url={https://openai.com/},
  author={openAI},
}

@article{doxygen,
  url={https://doxygen.nl/},
  author={Doxygen},
}

@article{svf,
  url={https://github.com/SVF-tools/SVF},
  author={SVF},
}

@article{bzip2,
  url={https://github.com/commontk/bzip2/blob/master},
  author={bzip2},
}

@article{libcsv,
  url={https://github.com/rgamble/libcsv/tree/master},
  author={libcsv},
}

@article{robotfindskitten,
  url={https://github.com/robotfindskitten/robotfindskitten},
  author={robotfindskitten},
}

@article{geiger,
  url={https://docs.rs/crate/cargo-geiger/latest},
  author={cargo-geiger},
}

@article{clippy,
  url={https://github.com/rust-lang/rust-clippy},
  author={Clippy},
}

@article{Claude,
  url={https://www.anthropic.com/index/introducing-claude},
  author={Claude},
}

@article{Gemini,
  url={https://blog.google/technology/ai/google-gemini-ai/},
  author={Gemini},
}

@article{KGC2Rust,
  url={https://github.com/FudanSELab/PtrTrans-C2Rust},
  author={KG-C2Rust},
}

@article{AFLplusplus,
  url={https://github.com/AFLplusplus/AFLplusplus},
  author={AFLplusplus},
}

@article{rosner2006wilcoxon,
  title={The Wilcoxon signed rank test for paired comparisons of clustered data},
  author={Rosner, Bernard and Glynn, Robert J and Lee, Mei-Ling T},
  journal={Biometrics},
  volume={62},
  number={1},
  pages={185--192},
  year={2006},
  publisher={Oxford University Press}
}

\end{document}